\documentclass[preprint,12pt]{elsarticle}

\usepackage{amssymb}
\usepackage{lineno,hyperref}
\usepackage{amsmath}
\usepackage{float}
\usepackage{subcaption}
\usepackage{mathtools, nccmath}
\usepackage[bb = px]{mathalfa}
\usepackage{bbm}
\usepackage{amssymb}
\usepackage{gensymb}
\usepackage{graphicx}
\usepackage{dcolumn}
\usepackage{rotating}
\usepackage{color}
\usepackage{times}
\usepackage{url,hyperref}
\usepackage{epstopdf,multirow}
\usepackage{geometry}
\usepackage{pstricks}
\usepackage{fancyhdr}
\usepackage{booktabs}
\usepackage{grffile}
\usepackage[font={small}]{caption}
\usepackage{subcaption}
\usepackage{multirow}
\usepackage{float}
\usepackage{sectsty}
\usepackage{siunitx}
\usepackage[]{lineno}
\usepackage[version=3]{mhchem}
\usepackage[section]{placeins}

\newcommand{\smallotimes}{\mathrel{\mkern-4mu\otimes\mkern-4mu}}

\newcommand{\intu}{\medint\int\displaylimits}
\DeclareMathOperator{\Tr}{Tr}

\makeatletter
\def\ps@pprintTitle{%
}
\makeatother

\begin{document}

\renewcommand{\arraystretch}{1.3}


%
%
%

\begin{frontmatter}
		
\title{A Hereditary Integral, Transient Network Approach to Modeling Permanent Set and Viscoelastic Response in Polymers}
\author[llnl]{Stephen T. Castonguay\corref{cor1}}
\author[cal,llnl]{Joshua B. Fernandes}
\author[llnl]{Michael A. Puso}
\author[llnl]{Sylvie Aubry}
\cortext[cor1]{Corresponding author.}
\address[llnl]{Lawrence Livermore National Laboratory, Livermore, CA 94551, USA}
\address[cal]{Department of Chemical and Biomolecular Engineering, University of California, Berkeley, Berkeley, CA 94720, USA}

\begin{abstract}
An efficient numerical framework is presented for modeling viscoelasticity and permanent set of polymers. It is based on the hereditary integral form of transient network theory, in which polymer chains belong to distinct networks each with different natural equilibrium states. Chains continually detach from previously formed networks and reattach to new networks in a state of zero stress. The free energy of these networks is given in terms of the deformation gradient relative to the configuration at which the network was born. A decomposition of the kernel for various free energies allows for a recurrence relationship to be established, bypassing the need to integrate over all time history. The technique is established for both highly compressible and nearly incompressible materials through the use of neo-Hookean, Blatz-Ko, Yeoh, and Ogden-Hill material models. Multiple examples are presented showing the ability to handle rate-dependent response and residual strains under complex loading histories. 
\end{abstract}



\end{frontmatter}

\section{Introduction}
Polymers put under mechanical loading often exhibit rate-dependence and/or permanent deformation tied to micro-scale configurational changes. This includes chain scission and additional crosslinking, both of which can lead to changes in the material stiffness. Furthermore, crosslinking in a strained deformation state can lead to a residual strain (or ``permanent set'') in the material. Additionally, the formation and breakage of weaker bonds among polymer chains also contributes to the macroscopic response, especially with respect to stress-strain rate dependence.

In the 1940s, Tobolsky and coworkers began studying stress relaxation and permanent set in polymeric materials \cite{tobolsky1944stress,tobolsky1945systems,andrews1946theory,green1946new}. Notably, Andrews, Tobolsky, and Hanson \cite{andrews1946theory} developed the ``two-network'' approach. They idealized the material as having separate independent networks, one in equilibrium in the unstrained reference state, and another in the strained deformation state. The proposition is that crosslinks associated with the original network break, causing stress relaxation, and new crosslinks then form in the strained state, which upon unloading would incur some residual deformation. For age-induced permanent set, the two-network framework has persisted, perhaps in part to the fact that the deformation state for many applications can be assumed static. For example, O-rings are often held in a constant strain state, simply undergoing stress relaxation. This model has therefore served as the basis for many chemical aging studies and models over the years \cite{zaghdoudi2019scission,rottach2004effect,rottach2006permanent,weisgraber2009mesoscopic,duncan2023improved,maiti2019age}.

Since their pioneering work, a plethora of models have been developed to describe the viscoelastic response of polymeric materials.
These generally fall into one of two frameworks: Hereditary Integral (HI) or Internal State Variable (ISV) approaches \cite{petiteau2013large,roylance2001engineering}. Internal variables represent phenomena not directly observed at the macroscopic level \cite{coleman1967thermodynamics,maugin2015saga}, and their use has since become commonplace when dealing with viscoelasticity. 
Finite linear viscoelasticity, which implies a linear kinetics model for the evolution of these variables, has become ubiquitous in finite element codes \cite{simo1987fully}. Extensions involving nonlinear evolution equations have been proposed in recent years as well \cite{reese1998theory,kumar2016two}. In hereditary integral approaches, the stress (or strain) is defined in terms of a functional of the material's history \cite{wineman2009nonlinear,drapaca2007nonlinear}. 
Phenomenological versions are generally defined in terms of the convolution of a relaxation function with some measure of stress or strain. In some cases, an equivalence can be made to internal variable approaches \cite{simo1987fully,simo2006computational}.

Transient network theory represents the class of models which generalize the two-network theory to continually deforming media, the earliest version of which was put forth in Green and Tobolsky's work on stress relaxation \cite{green1946new}. As with viscoelastic models in general, these approaches can be further classified into HI versus ISV approaches. For ISV versions, the state variables are associated with a distribution of chain end-to-end vectors \cite{tanaka1992viscoelastic,wang1992transient,vernerey2018transient, vernerey2017statistically}. Meanwhile HI approaches involve tracking the time history of chains detaching and attaching in an equilibrium state \cite{drozdov1999constitutive,long2014time,septanika1998application,shaw2005chemorheological}. An excellent comparison between the two is given by Hui et al. \cite{hui2021physically}.

In finite element modeling, the difficulty associated with the ISV approach to transient networks is the need to solve a Smoluchowski or Fokker-Planck equation for the chain distribution at each Gauss point, as well as the need to integrate over the configuration space to obtain the stress. Reduced versions can be obtained by assuming the Gaussian distribution chain model, eliminating the need to integrate over the configuration space, leading to a straightforward evolution equation for a configuration tensor \cite{linder2011micromechanically,vernerey2017statistically}. 
Meanwhile, the major drawback of the HI methods is simply the expense associated with the history integral. As a simulation proceeds, the number of past states needed for evaluation increases, leading to memory and computing requirements that increase linearly with the number of time steps.

While the state variable versions have proven more popular over the years, there have been several noteworthy contributions based on the HI form. Septanika et al. \cite{septanika1998application,septanika1998application2} and Budzien et al. \cite{budzien2008new} developed models based on the assumption that the material is held in a series of fixed strains, a direct extension of Tobolsky's two-network model. Drozdov published a number of HI models for both viscoelasticity and aging where the transient networks continually form and degrade \cite{drozdov1997constitutive,drozdov1998model,drozdov1999constitutive,drozdov1999model}. Hui and Long \cite{hui2012constitutive}, Long et al. \cite{long2014time}, and Guo  et al. \cite{guo2016mechanics} all developed models for self-healing gels in which a permanent network is used to model chemical bonds, and a transient network is used to model weaker physical bonds breaking and reforming. Other relevant models in the literature utilizing continuous networks include the works of Shaw et al. \cite{shaw2005chemorheological} and Ateshian \cite{ateshian2015viscoelasticity}.  

Due to the computational burden of HI methods, most of these models have only been employed numerically for simple deformations which are defined apriori, usually in the context of material testing. The only one that has been utilized in a finite element code for arbitrary deformations is Ateshian's model \cite{ateshian2015viscoelasticity} in the FEBio software \cite{maas2012febio}. Details on the implementation can be found in the recent paper from Ateshian et al. \cite{ateshian2023numerical}. In their numerical scheme, they employ a discrete network version, where rules for the merging of old networks and formation of new ones are introduced in an attempt to mitigate the total number of networks. However, this still entails storing and evaluating a number of past states, which can grow with time.

In this paper, we present a general form of transient network theory that is representative of each of the aforementioned HI models. We then proceed to demonstrate that under the assumption of first-order degradation kinetics, various forms of the network free energy permit a decomposition that allows for a recurrence relation to be established. This obviates the need to store and evaluate the history integral, leading to a simple and efficient implementation into finite element software.      

The outline of the paper is as follows: In Section \ref{section:HIF}, we present the model, show its thermodynamic consistency, and outline a recurrence relationship for material definitions which allow for the relevant decomposition. Section \ref{section:materials} then outlines the particular decompositions required for four popular material models: neo-Hookean, Blatz-Ko, Yeoh, and Ogden-Hill. Section \ref{sec:discuss} discusses potential advantages and disadvantages, and draws relations to other viscoelastic models in the literature. Additionally, representative examples are presented demonstrating stress relaxation and permanent set for both compressible foams and nearly incompressible polymers.

\section{Transient Network Theory} \label{section:HIF}
\subsection{Kinematics}
\begin{figure}
	\centering
	\includegraphics[trim={0cm 4cm 0cm 0cm},clip,width=0.9\textwidth]{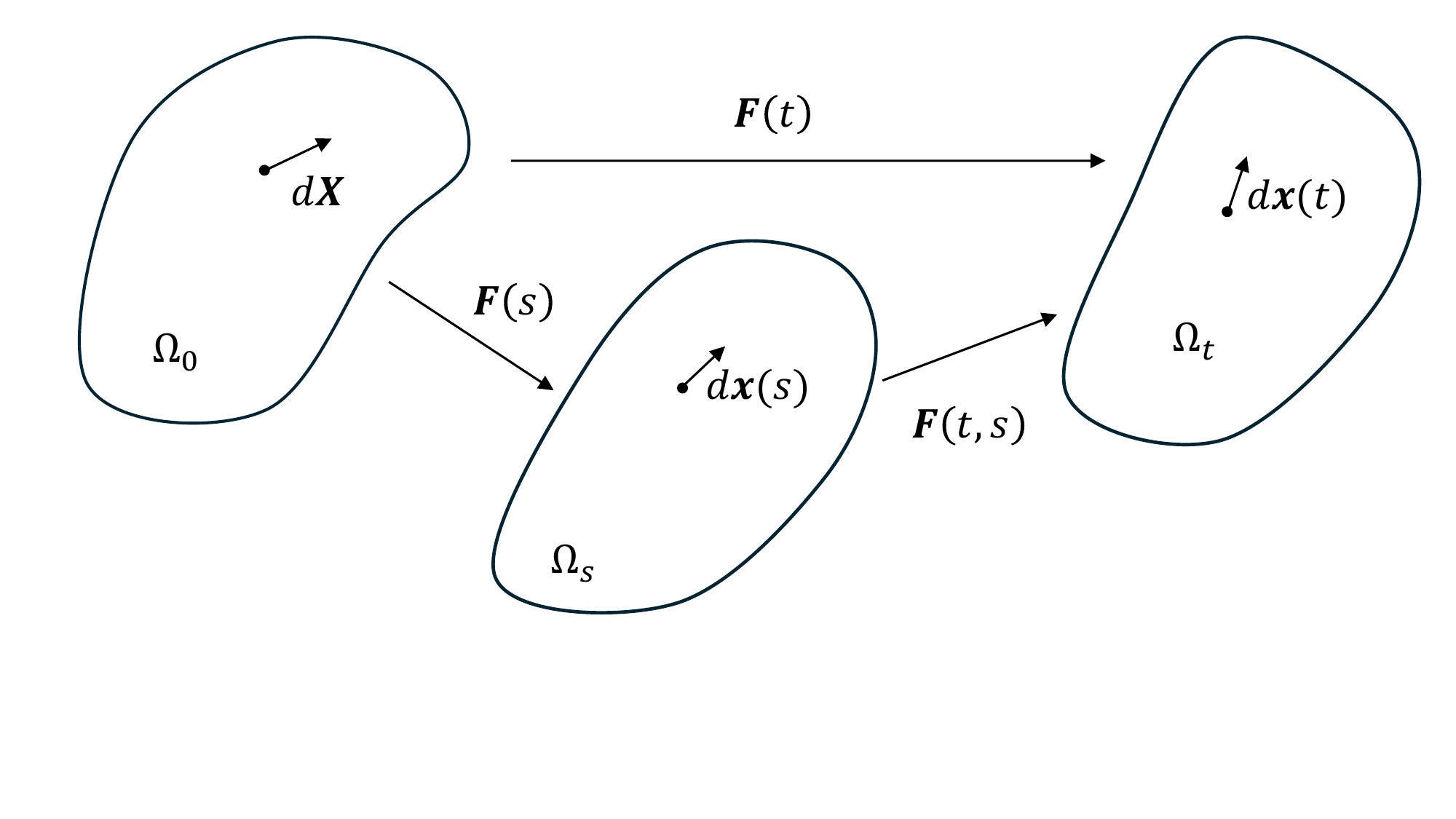}
   \caption{Infinitesimal line elements being transformed by deformation gradients relative to the original configuration and some past configuration at $s$}
\label{fig:RelConfigs}
\end{figure}

Let $\boldsymbol{X}$ describe the reference coordinates of a material point on the body, and $\boldsymbol{x}$ the current coordinates, both in 3D Euclidean space. Denote the domain of the body to be $\Omega_0$ and the boundary to be $\Gamma_0$ in the reference coordinates, and $\Omega_t$ and $\Gamma_t$ in the current coordinates.  The motion of the body is a one-parameter system (time $t$) given by the following mapping
\begin{equation}
	\boldsymbol{x}=\boldsymbol{\phi}(\boldsymbol{X},t) \, .
\end{equation}
The displacement of a material point is then
\begin{equation}
	\boldsymbol{u}(\boldsymbol{X},t)=\boldsymbol{\phi}(\boldsymbol{X},t)-\boldsymbol{X} \, ,
\end{equation}
while its velocity is given by
\begin{equation}\label{eqn:matlvel}
	\boldsymbol{v}=\dot{\boldsymbol{x}} = \frac{\partial \boldsymbol{\phi} (\boldsymbol{X},t)}{\partial t} \, ,
\end{equation}
where the dot represents the material time derivative. The deformation gradient is defined by
\begin{equation}\label{eqn:DefGrad}
	\boldsymbol{F}(\boldsymbol{X},t)=\nabla_X \boldsymbol{\phi}(\boldsymbol{X},t) \, .
\end{equation}
The material time derivative of the deformation gradient leads to the following identity 
\begin{equation} \label{eqn:velgraddef}
	\dot{\boldsymbol{F}} = \boldsymbol{L} \boldsymbol{F} \, ,
\end{equation}
where $\boldsymbol{L}=\nabla_x  (\boldsymbol{v}) $ is the velocity gradient. Additionally, the Jacobian of the deformation gradient, given by $J = \det \boldsymbol{F} $, relates the density in the current configuration $\rho$ to the reference density $\rho_0$ through
\begin{equation} \label{eqn:density}
	\rho_0(\boldsymbol{X}) = J \, \rho(\boldsymbol{x},t) \, .
\end{equation}
It will also be convenient to define motion relative to some past configuration at time $s$. In particular, the relative deformation gradient is given by
\begin{equation}\label{eqn:DefGradRel}
	\boldsymbol{F}(t,s)=\boldsymbol{F}(t)\boldsymbol{F}^{-1}(s) \, ,
\end{equation}
as depicted in Figure \ref{fig:RelConfigs}.
Throughout, $\boldsymbol{F}$ will refer to the deformation gradient relative to the original configuration (i.e.  $\boldsymbol{F}(t)=\boldsymbol{F}(t,0)$), unless stated otherwise.
\subsection{Balance Laws}
The finite element implementation used here involves an updated Lagrangian approach, so it is convenient to represent balance laws in the current configuration. Ignoring inertial terms, the force balance (equilibrium equation) is given by
\begin{equation}
 \nabla_x \cdot \boldsymbol{\sigma} =\boldsymbol{b} \, , 
\end{equation}
where $\boldsymbol{\sigma}$ is the Cauchy stress and $\boldsymbol{b}$ represents external body forces. Similarly, the energy balance is given by
\begin{equation}\label{eqn:EnergyBal}
	\begin{split}
		\rho \dot{e}=&\boldsymbol{\sigma} : \boldsymbol{L}  -\nabla_x \cdot (\boldsymbol{q}) +\rho \ s  \, ,
	\end{split}
\end{equation}
where $e$ is the specific internal energy, $\boldsymbol{q}$ is the heat flux, and $s$ represents external sources. 
The Clausius-Duhem inequality, which is a statement of the second law of Thermodynamics, can be expressed as
\begin{equation} \label{eqnCD_ineq}
	\rho \, \dot{\eta} \ge -\nabla_x \cdot (\frac{\boldsymbol{q}}{T}) +\frac{\rho s}{T} \, ,
\end{equation}
where $\eta$ is the specific entropy and $T$ is the temperature. 
Thermo-mechanical material models are often expressed through definitions of a Helmholtz free energy $\Psi$, which is related to the internal energy via the Legendre transform 
\begin{equation} \label{eqn:LegTran}
	e=\Psi+\eta T \, .
\end{equation}
Through Equations \ref{eqn:velgraddef}, \ref{eqn:EnergyBal}, and \ref{eqn:LegTran}, the Clausius-Duhem inequality may be expressed as
\begin{equation} \label{eqnCD_ineq2}
	- \rho \dot{\Psi} -\rho \eta \dot{T}  + (\boldsymbol{\sigma} \boldsymbol{F}^{-T}) : \dot{\boldsymbol{F}}   \ge \frac{1}{T}\nabla_x T \cdot \boldsymbol{q} \, .
\end{equation}
The heat flux is then assumed to obey $	\boldsymbol{q}=-\boldsymbol{k} \nabla_x T $, where $\boldsymbol{k}$ is a positive semi-definite conductivity tensor. Then, the second law is not violated if the following inequality is met
\begin{equation} \label{eqnCD_ineq3}
	- \rho \dot{\Psi} -\rho \eta \dot{T}  + (\boldsymbol{\sigma} \boldsymbol{F}^{-T}) : \dot{\boldsymbol{F}}   \ge 0 \, .
\end{equation} 
We also assume that the total specific free energy can be additively decomposed into thermal and mechanical parts as follows
\begin{equation} \label{eqn:sep_assumption}
	\Psi=\Psi_H(T(t)) +\Psi_M\big(* \big) \, ,
\end{equation}
where the thermal part is simply a function of the temperature $T$ at time $t$, while the mechanical part is a history functional which does not explicitly depend on the current temperature. Then, by using a Coleman-Noll argument \cite{coleman1967thermodynamics} for the temperature, $\eta =- \frac{\partial \Psi_H}{\partial T}$, and the inequality needed to satisfy the second law further reduces to 
\begin{equation} \label{eqnCD_ineq4}
	- \rho \dot{\Psi}_M  + (\boldsymbol{\sigma} \boldsymbol{F}^{-T}) : \dot{\boldsymbol{F}}   \ge 0 \, .
\end{equation} 
While we do not explicitly solve for the heat/energy equation in this work (and hence do not give a functional form for $\Psi_H$), the reason for assuming Equation \ref{eqn:sep_assumption} holds is that aging applications almost never occur under isothermal conditions, and this decomposition simplifies the derivation. 
\subsection{Hereditary Integral Formulation} \label{Sect:HIF}
A continuous extension of the two-network approach is given by the following functional representing the specific mechanical Helmholtz free energy  
\begin{equation} \label{eqn:FreeEnergy}
	\Psi_M\big(\xi(t,*),\boldsymbol{F}(t,*)\big)=\xi(t,0)\psi_C(\boldsymbol{F}(t))+\intu_{0}^{t}\xi(t,s)\psi_C(\boldsymbol{F}(t,s))ds \, ,
\end{equation}
where $\xi(t,s)$ is the density of chains (per unit mass) originated at time $s$ that remain intact at time $t$ and $\psi_C$ is the free energy per chain, which is a function of the relative deformation gradient. 
Since $\xi(t,s)$ represents the continual creation of networks, it is a quantity per unit mass and time.  $\xi(t,0)$, on the other hand, represents the original network, and is simply a quantity per unit mass \footnote{Some authors (e.g. \cite{drozdov1997constitutive,long2014time}) formulate the free energy through a variable $\hat{\xi}(t,s)$ defined by $\frac{\partial \hat{\xi}(t,s)}{\partial s}:=\xi(t,s)$, such that $\hat{\xi}(t,s)$ and $\xi(t,0)$ have the same units.}.
Chain densities in this context associate parts of the material with distinct natural equilibrium states. 
Now, taking the time derivative of Equation \ref{eqn:FreeEnergy} leads to
\begin{equation} \label{eqn:DeriveMechEnergy}
	\begin{split}
			\dot{\Psi}_M(t)= &	\dot{\xi}(t,0)\psi_C(\boldsymbol{F}(t))+	\xi(t,0)\frac{\partial \psi_C(\boldsymbol{F}(t))}{\partial \boldsymbol{F}(t)}	:\dot{\boldsymbol{F}}(t)\\
			&+\intu_{0}^{t}\dot{\xi}(t,s)\psi_C(\boldsymbol{F}(t,s)) + \xi(t,s)\frac{\partial \psi_C(\boldsymbol{F}(t,s))}{\partial \boldsymbol{F}(t,s)}:\boldsymbol{F}^{-1}(s)\dot{\boldsymbol{F}}(t) ds \, ,
	\end{split}
\end{equation}
where we have used the Leibniz rule, as well as the condition that $\psi_C(\boldsymbol{F}(t,t))=0$. 
Inserting Equation \ref{eqn:DeriveMechEnergy} into \ref{eqnCD_ineq4} and applying a Coleman-Noll type argument for the deformation gradient leads to 
\begin{equation} \label{eqn:CauchyStress}
	\boldsymbol{\sigma} = \rho \xi(t,0)\frac{\partial \psi_C(\boldsymbol{F}(t))}{\partial \boldsymbol{F}(t)} \boldsymbol{F}^{T}(t)+ \rho \intu_{0}^{t}  \xi(t,s)\frac{\partial \psi_C(\boldsymbol{F}(t,s))}{\partial \boldsymbol{F}(t,s)} \boldsymbol{F}^{-T}(s)\boldsymbol{F}^{T}(t)  ds \, .
\end{equation}
The second law then is further reduced to
\begin{equation}
	\dot{\xi}(t,0)\psi_C(\boldsymbol{F}(t))	+\intu_{0}^{t}\dot{\xi}(t,s)\psi_C(\boldsymbol{F}(t,s))ds \le 0 \, .
\end{equation}
Networks form instantaneously in a state of zero energy and then subsequently decay, so that 
\begin{align}
	\dot{\xi}(t,s) \ge 0  ; \, \psi_C(\boldsymbol{F}(t,s)) =0   \, ,  &\quad \quad t = s \\
	\dot{\xi}(t,s) \le 0 ; \, \psi_C(\boldsymbol{F}(t,s)) \ge 0  \, ,  & \quad \quad t > s \, .
\end{align}
Therefore, the second law is explicitly satisfied.

\subsection{Network Kinetics}
The degradation kinetics are assumed to be first-order, and therefore are given by
\begin{equation} \label{eqn:kinetics}
	\dot{\xi}(t,s)=-k(t)  \xi(t,s) \quad \forall    t > s  \, ,
\end{equation}
where $k(t)$ represents the degradation rate. 
Through stress relaxation experiments on various rubbers, Tobolsky found that this rate is deformation independent (at least up to stretch values of about 4), but strongly dependent on temperature \cite{tobolsky1944stress}. This assumption was further verified by the works of Scanlon and Watson \cite{scanlan1958interpretation} and Jones \cite{jones2003experimental}. 
Experimental evidence has also been shown that the kinetics of secondary networks associated with physical bonds in gels are also deformation independent \cite{mayumi2013stress}. 
Therefore, any dependence on time is specifically through the temperature (i.e. $k=k(T(t))$.

Equation \ref{eqn:kinetics} has the following general solution
\begin{equation}
	\xi(t,s) =G(s)\exp\Big( -\intu_s^t k(\tau) d\tau \Big) \, .
\end{equation}
Here, $G$ represents the generation of a network associated with chain attachment at a particular point in time. 
We make the assumption that polymer chains breaking from their networks immediately reform in a new network, so that the total number of effective chains over all networks is constant. Therefore, the following constraint on the total network density is imposed
\begin{equation} \label{eqn:chainconstraint}
	\xi_0-\xi(t,0)=\intu_{0}^t \xi(t,s) ds \, ,
\end{equation}
where $\xi_0:=\xi(0,0)$. This states that the number of chains that have detached from the original network must equal the number currently attached to the continually forming networks.
The following relation is then obtained by taking the time derivative of Equation \ref{eqn:chainconstraint}
\begin{equation}\label{eqn:cc_aux}
	-\dot{\xi}(t,0)=\xi(t,t)+\intu_{0}^t \dot{\xi}(t,s) ds \, .
\end{equation}
Inserting Equations \ref{eqn:kinetics} and \ref{eqn:chainconstraint} into \ref{eqn:cc_aux} and using $\xi(t,t) =G(t)$ leads to $ G(t)= k(t)\xi_0 $. 
Therefore, the chain density may be written as 
\begin{equation} \label{eqn:ChainSoln}
	\xi(t,s) =k(s)\xi_0\exp\Big( -\intu_s^t k(\tau) d\tau \Big) \, .
\end{equation}
Now, in addition to defining a degradation rate $k$, the model requires a definition of the chain free energy $\psi_C$ and an initial condition for the chain density $\xi_0$. However, one can simply define a macroscopic free energy $\psi=\xi_0\psi_C$ and use the chain fraction, defined by
\begin{equation} \label{eqn:ChainFrac}
	 \gamma(t,s):=\frac{\xi(t,s)}{\xi_0} \, .
\end{equation}
This can be used directly in the definition for the Helmholtz free energy, however the form given in Equation \ref{eqn:FreeEnergy} is presented to connect with similar derivations in the literature (e.g. \cite{drozdov1999constitutive} \cite{hui2021physically}).

While the framework outlined in the Section \ref{Sect:HIF} is representative of a variety of HI transient network models, there are various alternatives to how the formation and degradation of networks is handled. For instance, one may simply specify chain breakage and growth functions directly. If these are not constrained by a constant chain density, then the material can stiffen or degrade, often associated with aging of a material  \cite{lion2012representation,septanika1998application2}. 
Another modeling choice for network evolution is given by Long et al. \cite{long2014time} for the behavior of self-healing gels. 
In their work, there are a fixed total number of bonds, but they assume the reformation of bonds in a new network is not immediate. Instead, the network formation process follows its own evolution equation in which the rate is proportional to the number of available broken bonds. 

The numerical procedures and recurrence relations outlined in the following sections are restricted to cases where the degradation of continually forming networks is governed by first-order kinetics (or equivalently, by exponential decay). However, no such restrictions are placed on the generation of new networks. The constant chain and immediate reformation assumptions, which govern the formation of new networks, are therefore chosen for simplicity in this work.

\subsection{Numerical Procedure} \label{sec:NumericalProc}
\subsubsection{Transient Network Stress Representation}
Using the chain fraction defined in Equation \ref{eqn:ChainFrac}, the stress from all the induced networks is given by
\begin{equation}
	\boldsymbol{\sigma}^{**}(t) = \intu_{0}^{t}  \gamma(t,s)\boldsymbol{\sigma}^*(t,s)ds
\end{equation}
where the stress component of a particular transient network is defined by 
\begin{equation}
	\boldsymbol{\sigma}^*(t,s) :=\rho \frac{\partial \psi(\boldsymbol{F}(t,s))}{\partial \boldsymbol{F}(t,s)}\boldsymbol{F}^T(t,s) \, .
\end{equation}
Then we aim to decompose $\boldsymbol{\sigma}^*$ into a general split

\begin{equation} \label{decomp1}
		\boldsymbol{\sigma}^*(t,s)=\sum_{i}\boldsymbol{A}^i(s) : \boldsymbol{B}^i(t)  \, ,
\end{equation}
where $\boldsymbol{A}$ and $\boldsymbol{B}$ may represent scalars or higher rank tensors \footnote{To disambiguate the notation for arbitrary rank tensors in Equation \ref{decomp1}, the $:$ operator (in an abuse of notation) represents a contraction of the form $A_{MN...}(s) B_{MN...ij}(t)$, where Cartesian index notation is used.}. Note that this is not always possible or feasible, depending on the choice of the free energy. However, if it is, then we are able to express the stress through 

\begin{equation} \label{StressDef}
	\begin{split}
		\boldsymbol{\sigma}^{**}(t) &= \sum_{i}\boldsymbol{\mathcal{H}}^i(t) : \boldsymbol{B}^i(t) \, ,
	\end{split}
\end{equation}
where $\boldsymbol{\mathcal{H}}^i$ represents history variables defined by
\begin{equation} \label{HistDef}
	\begin{split}
		\boldsymbol{\mathcal{H}}^i(t) &=\intu_{0}^{t}  \gamma(t,s)\boldsymbol{A}^i(s)ds \, .
	\end{split}
\end{equation}
A corresponding rate form for the evolution of these history variables can be derived by taking the time derivative of Equation \ref{HistDef}, leading to
\begin{equation}\label{rateform1}
	\begin{split}
		\dot{\boldsymbol{\mathcal{H}}}^i(t) &= \gamma(t,t)\boldsymbol{A}^i(t)+\intu_{0}^{t}  \dot{\gamma}(t,s)\boldsymbol{A}^i(s)ds \, .
	\end{split}
\end{equation}
Through Equations \ref{eqn:ChainSoln} and \ref{eqn:ChainFrac}, $\gamma(t,t)=k(t)$, and by further utilizing Equations \ref{eqn:kinetics} and \ref{HistDef}, Equation \ref{rateform1} may be re-written as
\begin{equation}\label{rateformf}
	\begin{split}
		\dot{\boldsymbol{\mathcal{H}}}^i(t) =k(t) \Big( \boldsymbol{A}^i(t) -\boldsymbol{\mathcal{H}}^i(t)\Big) \, .
	\end{split}
\end{equation}
In the appendix of Green and Tobolsky's pioneering work \cite{green1946new}, they derived the above equation for the particular case of an incompressible neo-Hookean model, where $\boldsymbol{A}(t)=\boldsymbol{C}^{-1}(t)$ and $\boldsymbol{\boldsymbol{\mathcal{H}}}$ is a second rank history tensor.

\subsubsection{Integration of the History Variable}
For the development of a numerical method, the solution at $t_{N+1}=t_N+\Delta t$ is sought, given information at $t_N$. The chain density can be updated straightforwardly via
\begin{equation}
	\gamma(t_{N+1},s) =k(s)\exp\big( -\intu_s^{t_{N+1}} k(\tau) d\tau \big) =\gamma(t_N,s)\exp\big( -\intu_{t_N}^{t_{N+1}} k(\tau) d\tau \big) \, .
\end{equation}
This can be used to update the history variables through
\begin{equation}\label{eqn:GenRecurrence}
	\begin{split}
		\boldsymbol{\mathcal{H}}^i (t_{N+1}) =&\exp\big( \intu_{t_N}^{t_{N+1}} -k(\tau) d\tau \Big)\intu_{0}^{t_N}  \gamma(t_N,s)\boldsymbol{A}^i(s)ds +\intu_{t_N}^{t_{N+1}}  \gamma(t_{N+1},s)\boldsymbol{A}^i(s)ds  \\
		=& \exp\big( \intu_{t_N}^{t_{N+1}} -k(\tau) d\tau \big)\boldsymbol{\mathcal{H}}^i(t_N) +\intu_{t_N}^{t_{N+1}}  \gamma(t_{N+1},s)\boldsymbol{A}^i(s)ds  \, .
	\end{split}
\end{equation}
This represents a general recurrence relationship and can be treated in a number of different ways. For instance, one could numerically integrate Equation \ref{eqn:GenRecurrence} or use a finite difference scheme on the equivalent rate form of Equation \ref{rateformf}. For simplicity, we take $k$ and $\boldsymbol{A}$ to be constant through the time step, denoted by $\tilde{k}$ and $\boldsymbol{\tilde{A}}$, so that it can be analytically integrated as follows
\begin{equation}
	\begin{split}
		\boldsymbol{\mathcal{H}}^i(t_{N+1}) = \exp(-\tilde{k}\Delta t )\boldsymbol{\boldsymbol{\mathcal{H}}}^i(t_N) +\boldsymbol{\tilde{A}}^i  \Big(1 - \exp(-\tilde{k} \Delta t)  \Big)  \, .
	\end{split}
\end{equation}
The constants $\boldsymbol{\tilde{A}}$ and $\tilde{k}$ are then approximated by the average of endpoint values (i.e. $\boldsymbol{\tilde{A}}=\frac{\boldsymbol{A}_{N+1}+\boldsymbol{A}_N}{2}$).

\section{Material Models} \label{section:materials}
Finite strain material models are often defined in terms of the referential volumetric free energy density $\tilde{W}(\boldsymbol{F})=\rho_s \psi(\boldsymbol{F})$, where here the reference configuration is at time $s$.
Additionally, for isotropic materials, the functional form is usually specified in terms of the right Cauchy-Green tensor $\boldsymbol{C}=\boldsymbol{F}^T \boldsymbol{F}$ such that  $\tilde{W}(\boldsymbol{F})=W(\boldsymbol{C})$.  This leads to straightforward expressions for the second Piola-Kirchhoff stress
\begin{equation}
	S^{*}_{IJ}= 2  \frac{\partial W(\boldsymbol{C})}{\partial C_{IJ}} \, ,
\end{equation}
where we have introduced Cartesian index notation, which will be convenient for expressing tensors associated with Equation \ref{decomp1}.
Additionally, the material tangent is defined by 
\begin{equation}
	H^{*}_{IJKL}=  4 \frac{\partial^2 W}{\partial C_{IJ}\partial C_{KL}} = 2  \frac{\partial S^*_{IJ}}{\partial C_{KL}} \, .
\end{equation}
The Cauchy stress and spatial tangent are then found through Piola pushforwards of these quantities, as follows
\begin{equation} \label{Piola_push1}
	\sigma^{*}_{ij}= \frac{1}{J}  F_{iM} S^{*}_{MN} F_{jN}
\end{equation}
\begin{equation}\label{Piola_push2}
	H^{*}_{ijkl}= \frac{1}{J}  F_{iM} F_{jN} 	H^{*}_{MNOP} F_{kO} F_{lP} \, .
\end{equation}
For the transient networks, these expressions are evaluated with respect to the relative deformation gradient. Decompositions of each may then be found, if possible, as demonstrated in the following sections for the stress, and in \ref{section:Tangents} for the tangent. 
\subsection{Invariant-Based Models for Highly Compressible Materials} \label{section:highcomp_inv_materials}
\subsubsection{Compressible Neo-Hookean}
To maintain frame invariance, many materials models are based on the following principal invariants of the right Cauchy-Green tensor

\begin{equation}
	I_1=\Tr\big(\boldsymbol{C}\big)=C_{II}
\end{equation}

\begin{equation}
	I_2=\frac{1}{2}\Big(\big(\Tr\big(\boldsymbol{C}\big)\big)^2 - \Tr\big(\boldsymbol{C}^2\big) \Big)=\frac{1}{2}\Big(C_{II}C_{JJ}-C_{IJ}C_{JI} \Big)
\end{equation}

\begin{equation}
	I_3=\det(\boldsymbol{C})=J^2 \, .
\end{equation}
For incompressible materials, $I_3=1$, and the neo-Hookean model takes its ubiquitous form
\begin{equation}
	W= \frac{\mu}{2}(I_1-3) \, ,
\end{equation}
where $\mu$ is Lam\'{e}'s second parameter (shear modulus). However, there exists various extensions for compressible materials as well \cite{kossa2023analysis}. One particularly suited for highly compressible materials is given by \cite{bonet1997nonlinear}
\begin{equation}
	W= \frac{\mu}{2}(I_1-3 -2\ln(J)) +  \frac{\lambda}{2} (\ln(J))^2 \, ,
\end{equation}
where $\lambda$ is the first Lam\'{e} parameter. 
The Cauchy stress associated with this functional form is then
\begin{equation}
	\begin{split}
		\sigma_{ij} = \frac{\mu}{J}   (b_{ij}-\delta_{ij}) +\frac{\lambda}{J} \ln(J) \delta_{ij} \, ,
	\end{split}
\end{equation}
where $\boldsymbol{b}=\boldsymbol{F}\boldsymbol{F}^T$ is the left Cauchy-Green tensor and  $\boldsymbol{\delta}$ is the identity tensor.
By inserting the corresponding expressions for the relative deformation gradient (see \ref{section:RelIdentities}), the stress is decomposed via the terms in Table \ref{CNHDecomp}.

\begin{table}
		\centering
		\begin{tabular}{| r | l || r | l | }
			\hline 
			$A^{1}_{MN} $ &$J(s)C_{MN}^{-1}(s) $ &$B^{1}_{MNij}  $ &$ \frac{\mu}{J(t)}  F_{iM}(t)F_{jN}(t)$     \\ 
			$A^{2} $ & $J(s) $ &$B^{2}_{ij} $ & $  -\frac{\mu}{J(t)}  \delta_{ij}+\frac{\lambda\ln(J(t))}{J(t)}  \delta_{ij} $ \\
			$A^{3}$&$J(s)\ln(J(s))$ &$B^{3}_{ij} $&$		-\frac{\lambda}{J(t)}  \delta_{ij}$ \\
			\hline 
		\end{tabular}
	\caption{Compressible neo-Hookean decomposition}
	\label{CNHDecomp}
\end{table}

\subsubsection{Blatz Ko}
While many material models associated with polymers have been developed under the assumption of incompressibility, Blatz and Ko \cite{blatz1962application} performed experiments and developed a model for highly compressible elastomeric foams. Notably, they introduce a different set of invariants for their functional form, as follows
\begin{equation}
	J_1=I_1=C_{II}
\end{equation}
\begin{equation}
	\begin{split}
		J_2=\frac{I_2}{I_3}=\frac{1}{2J^2}\Big(C_{II}C_{JJ}-C_{IJ}C_{JI} \Big)=(C^{-1})_{II}
	\end{split}
\end{equation}
\begin{equation}
	\begin{split}
		J_3=\sqrt{I_3}=\det(F)=J \, .
	\end{split}
\end{equation}
The particular strain energy density and Cauchy stress of their model is given by
\begin{equation} \label{BlatzKoSED}
	W = f \frac{\mu}{2} \Big(J_1-3 +\frac{2}{\beta}( J^{-\beta}_3 -1) \Big)+ (1-f)\frac{\mu}{2} \Big(J_2-3 +\frac{2}{\beta}( J^{\beta}_3 -1) \Big)
\end{equation} 

\begin{equation} \label{BlatzKoStress}
	\begin{split}
		\sigma_{ij} = \frac{1}{J} f   \mu b_{ij} - \frac{1}{J}(1-f) \mu b^{-1}_{ij} - \frac{1}{J}\mu \Big( f J^{-\beta} - (1-f) J^{\beta} \Big) \delta_{ij}  \, .
	\end{split}
\end{equation} 
In this expression, $\beta=\frac{2\nu}{1-2\nu}$, where $\nu$ is Poisson's ratio. Taking $f$ to be one reduces to a compressible neo-Hookean material, a form Blatz and Ko used to characterize rubbers. They also took $f$ to be zero and $\nu$ to be $\frac{1}{4}$ to model foams. This latter case is often referred to as the specialized Blatz Ko model, which has been implemented in a variety of commercial codes (e.g. \cite{systemes2011solidworks,thompson2017ansys}).  Table \ref{BKDecomp} outlines the decomposition for the transient network stresses for the more general form given by Equations \ref{BlatzKoSED} and \ref{BlatzKoStress}.

\begin{table}[h]
	\centering
	\begin{tabular}{| r | l || r | l | } 
		\hline
		$A^{1}_{MN} $ &$J(s)C_{MN}^{-1}(s) $ &$B^{1}_{MNij} $&$ f\frac{\mu}{J(t)}  F_{iM}(t)F_{jN}(t)$   \\ 
	$A^{2}_{MN} $ &$J(s)C_{MN}(s) $  &$B^{2}_{MNij} $&$ -(1-f) \frac{\mu}{J(t)}  F^{-1}_{Mi}(t)F^{-1}_{Nj}(t)$  \\
	$A^{3} $ & $(J(s))^{1+\beta} $ &$B^{3}_{ij} $&$-f\mu \big(J(t)\big)^{-\beta-1}  \delta_{ij}$ \\
	$A^{4} $ & $(J(s))^{1-\beta} $ 	&		$B^{4}_{ij} $&$(1-f)\mu \big(J(t)\big)^{\beta-1}  \delta_{ij}$ \\
		\hline 
	\end{tabular}
	\caption{Blatz-Ko decomposition}
\end{table} \label{BKDecomp}

\subsection{Stretch-Based Material Models} \label{section:stretch_materials}
Models based on principal stretches are the primary alternative to invariant-based material models. This concept starts with a polar decomposition of the deformation gradient through $\boldsymbol{F}=\boldsymbol{R} \boldsymbol{U}=\boldsymbol{V} \boldsymbol{R}$, where $\boldsymbol{R}$ is a rotation tensor and $\boldsymbol{U}$ and $\boldsymbol{V}$ are the right and left stretch tensors, respectively. The spectral representations of $\boldsymbol{U}$ and $\boldsymbol{V}$ are then given by
\begin{equation}
	\boldsymbol{U}= \sum_{a=1}^{3} \lambda_a \boldsymbol{N}_a \otimes \boldsymbol{N}_a = \sum_{a=1}^{3} \lambda_a \boldsymbol{M}_a 
\end{equation}
\begin{equation}
	\boldsymbol{V}= \sum_{a=1}^{3} \lambda_a \boldsymbol{n}_a \otimes \boldsymbol{n}_a = \sum_{a=1}^{3} \lambda_a \boldsymbol{m}_a \, .
\end{equation}
Here, $\lambda_a$ are the principal stretches, $\boldsymbol{N}_a$ and $\boldsymbol{n}_a$ represent the orthonormal eigenvectors of the stretch tensors, and $\boldsymbol{M}_a$ and $\boldsymbol{m}_a$ represent the corresponding eigenbases. 
The Cauchy-Green deformation tensors are related to the stretches as follows
\begin{equation}
	\boldsymbol{C}=\boldsymbol{U}\boldsymbol{U}=  \sum_{a=1}^{3} \lambda_a^2 \boldsymbol{M}_a ; \quad \quad
	\boldsymbol{b}=\boldsymbol{V}\boldsymbol{V}=  \sum_{a=1}^{3} \lambda_a^2 \boldsymbol{m}_a \, .
\end{equation}
One of the most popular stretch-based models for foams, known as the Ogden-Hill model \cite{storaakers1986material,ogden1997non,hill1979aspects}, is given by
\begin{equation} \label{Ogden1}
	\tilde{W}(\boldsymbol{U}) = \sum_{p=1}^{N} \frac{\mu_p}{\tilde{\alpha}_p} \Big(\lambda_1^{\tilde{\alpha}_p}+\lambda_2^{\tilde{\alpha}_p}+\lambda_3^{\tilde{\alpha}_p}-3 + \frac{1}{\beta}(J^{-\tilde{\alpha}_p \beta_p}-1)  \Big) \, .
\end{equation}
By defining $\tilde{\alpha}_p=2\alpha_p$, this can also be expressed as
\begin{equation} \label{Ogden2}
	W(\boldsymbol{C}) =W(\boldsymbol{U}^2) = \sum_{p=1}^{N} \frac{\mu_p}{2\alpha_p} \Big(\lambda_1^{2\alpha_p}+\lambda_2^{2\alpha_p}+\lambda_3^{2\alpha_p}-3 + \frac{1}{\beta_p}(J^{-2\alpha_p \beta_p}-1)  \Big) \, .
\end{equation}
Because the eigenbasis will be different for each relative deformation gradient, we turn to the concept of generalized strains, which contain information about the stretch eigenbasis. Various strain generalizations have been proposed in the literature, but the one most relevant here is the Seth-Hill family of strains \cite{Seth1962,Hill1968}, given by
\begin{equation}
	\begin{split}
		\boldsymbol{E}^{\alpha}&:=\frac{1}{2\alpha}\Big(\boldsymbol{U}^{2\alpha} -\boldsymbol{I}\Big)  \, \, \, \, \, \, \, \alpha\ne 0 \\
		\boldsymbol{E}^{0}&:=\ln \big(\boldsymbol{U} \big) 
	\end{split}
\end{equation}
where
\begin{equation}
	\begin{split}
		\boldsymbol{U}^{2\alpha}&= \sum_{a=1}^{3} \lambda^{2\alpha}_a \boldsymbol{M}_a  \quad \alpha\ne 0 \\
		\ln\big(\boldsymbol{U}\big)&=\sum_{a=1}^{3} \ln(\lambda_a) \boldsymbol{M}_a \, . 
	\end{split}	
\end{equation}
This family of strains encompasses many popular strain measures, including Almansi ($\alpha=-1$), Hencky ($\alpha=0$), Biot ($\alpha=\frac{1}{2}$),
and Green ($\alpha=1$) strains.
We see that we can rewrite Equation \ref{Ogden1} through invariants of the Seth-Hill stretches as follows 
\begin{equation} \label{OgdenAlt}
	W(\boldsymbol{U}^2) = \sum_{p=1}^{N} \frac{\mu_p}{2\alpha_p} \Big( \Tr\big(\boldsymbol{U}^{2\alpha_p}\big)-3 + \frac{1}{\beta_p}(\det \big(\boldsymbol{U}^{2\alpha_p}\big)^{-\beta_p}-1)  \Big) \, .
\end{equation}
The relative deformation gradient can be written as $\boldsymbol{F}(t,s)=\boldsymbol{F}(t)\boldsymbol{F}^{-1}(s)= \boldsymbol{R}(t) \boldsymbol{\check{U}}(t,s)\boldsymbol{R}(s)$, where $\boldsymbol{\check{U}}(t,s):=\boldsymbol{U}(t) \boldsymbol{U}^{-1}(s)$ is a relative stretch tensor. 
A generalized version of this relative stretch tensor may then be defined through $\boldsymbol{\check{U}}^{2\alpha}(t,s):=\boldsymbol{U}^{2\alpha}(t) \boldsymbol{U}^{-2\alpha}(s)$. Note that this definition differs from the generalized stretch $\boldsymbol{U}^{2\alpha}(t,s)$ obtained via $ \boldsymbol{F}(t,s)=\boldsymbol{R}(t,s)\boldsymbol{U}(t,s)$.
Now, inserting $\boldsymbol{\check{U}}^{2\alpha}(t,s)$ into Equation \ref{OgdenAlt} leads to
\begin{equation} \label{BlatzKoSED}
	W(\boldsymbol{U}^2(t),\boldsymbol{U}^2(s)) = \sum_{p=1}^{N} \frac{\mu_p}{2\alpha_p} \Big( \boldsymbol{U}^{2\alpha_p}(t)  : \boldsymbol{U}^{-2\alpha_p}(s)   -3 + \frac{1}{\beta_p}\big( (\frac{J(t)}{J(s)})^{-2\alpha_p\beta_p}-1\big) \Big) \, ,
\end{equation}
where symmetry of the generalized stretches has been utilized. 
Unlike the invariant-based models, the free energy is not defined directly in terms of the relative deformation gradient. To facilitate the derivation of the kernel decomposition, we 
rewrite Equation \ref{eqn:CauchyStress} directly in terms of $\boldsymbol{F}(t)$
as follows
\begin{equation}
	\boldsymbol{\sigma} = \gamma(t,0)\frac{1}{J(t)} \frac{\partial \tilde{W}(\boldsymbol{U}(t))}{\partial \boldsymbol{F}(t)} \boldsymbol{F}^{T}(t)+  \intu_{0}^{t}  \gamma(t,s)\frac{J(s)}{J(t)}\frac{\partial \tilde{W}(\boldsymbol{U}(t),\boldsymbol{U}(s))}{\partial \boldsymbol{F}(t)} \boldsymbol{F}^{T}(t)  ds \, .
\end{equation}
Employing $\frac{\partial \tilde{W}(\boldsymbol{U}(t))}{\partial \boldsymbol{F}(t)}=2\boldsymbol{F}\frac{\partial W(\boldsymbol{U}^2(t))}{\partial \boldsymbol{U}^2(t)}$, the Cauchy stress associated with the induced networks can be expressed through the Piola pushforward given by Equation \ref{Piola_push1}, where the second Piola-Kirchhoff stress is given by

\begin{equation}
	\boldsymbol{S}^{**}(t) = \intu_{0}^{t}  \gamma(t,s)\boldsymbol{S}^*(t,s)  ds ; \quad 	\boldsymbol{S}^*(t,s) = 2J(s)\frac{\partial W(\boldsymbol{U}^2(t),\boldsymbol{U}^2(s))}{\partial \boldsymbol{C}(t)} \, .
\end{equation}
A decomposition is then sought of the form 
$\boldsymbol{S}^*(t,s)  =\sum_{i}\boldsymbol{A}^i(s) : \boldsymbol{B^*}^i(t) $, so that the Cauchy stress is given by Equation \ref{StressDef}, with $\boldsymbol{B}^i(t) =\frac{1}{J(t)}\boldsymbol{F}(t) \boldsymbol{B^*}^i(t) \boldsymbol{F}^{T}(t)$.
For the term associated with the trace of a particular stretch measure
%
%

\begin{equation}
	S^{*}_{IJ}=  \mu  J(s)  U_{MN}^{-2\alpha}(s) P^{\alpha}_{MNIJ}(t) \, ,
\end{equation}
where 
\begin{equation}
	P^{\alpha}_{MNIJ}=2 \frac{\partial E_{MN}^{\alpha}}{\partial C_{IJ}} \, .
\end{equation}
The derivation of the tensor $\boldsymbol{P}^{\alpha}$ and its derivative with respect to $\boldsymbol{C}$ (required for the tangent, see \ref{section:Tangents}) can be found in other works utilizing generalized strains (e.g. \cite{miehe2001algorithms, liu2024continuum}). 
For the case of three distinct eigenvalues, the principal directions are unique and $\boldsymbol{P}^{\alpha}$ may be expressed as
\begin{equation}
	\boldsymbol{P}^{\alpha}=  \sum_{a=1}^{3} d_a \boldsymbol{M}_a \otimes \boldsymbol{M}_a +\sum_{a=1}^{3}\sum_{b\ne a}^{3} v_{ab} \boldsymbol{G}^*_{ab} \, ,
\end{equation}
with
\begin{equation}
	\boldsymbol{G}^*_{ab}:=   \boldsymbol{N}_a \otimes  \boldsymbol{N}_b  \otimes  \boldsymbol{N}_a  \otimes  \boldsymbol{N}_b +\boldsymbol{N}_a \otimes  \boldsymbol{N}_b  \otimes  \boldsymbol{N}_b  \otimes  \boldsymbol{N}_a 
\end{equation}
\begin{equation}
	d_a := 2 \frac{\partial E_a^{\alpha}}{\partial \lambda_a^2}=  \lambda_a^{2(\alpha-1)}
\end{equation}
\begin{equation}
	v_{ab}:= \frac{E_a^{\alpha} (\lambda_a)-E_b^{\alpha} (\lambda_b)}{\lambda^2_a-\lambda^2_b} =\frac{1}{2\alpha} \frac{\lambda_a^{2\alpha} -\lambda_b^{2\alpha} }{\lambda^2_a-\lambda^2_b} \, ,
\end{equation}
where $E_a^{\alpha}(\lambda_a)$ is defined through $\boldsymbol{E}^{\alpha} = \sum_{a=1}^3 E^{\alpha}_a (\lambda_a) \boldsymbol{M}_a$. In the case of repeated eigenvalues
\begin{equation}
	\lim\limits_{\lambda_b \rightarrow \lambda_a} v_{ab}=\frac{1}{2}d_a \, .
\end{equation}

Unlike the trace term, the term associated with the determinant of a particular stretch measure is expressed directly in terms of the third invariant and does not include the spectral basis. This term has been presented as part of the Blatz-Ko material model and can be treated the same way. In fact, the Blatz-Ko model is a special case of the Ogden-Hill model, where $\alpha_1=1$ and $\alpha_2=-1$. The decomposition for the terms associated with the Ogden-Hill Cauchy stress is given in Table \ref{OgdenDecomp}.

\begin{table}[h] 
	\centering
	\begin{tabular}{| r | l || r | l | } 
		\hline
		$A^{\Tr}_{MN} $ &$J(s)  U_{MN}^{-2\alpha}(s)$ &$B^{\Tr}_{MNij} $&$ \frac{\mu}{J(t)}  F_{iK}(t)P^{\alpha}_{MNKL}(t)F_{jL}(t)$   \\ 
		$A^{\det} $ & $(J(s))^{2\alpha\beta+1} $ &$B^{\det}_{ij} $&$-\mu \big(J(t)\big)^{-2\alpha\beta-1}  \delta_{ij}$ \\
		\hline 
	\end{tabular}
	\caption{Ogden-Hill decomposition}
	\label{OgdenDecomp}
\end{table}

\subsection{Nearly Incompressible Materials} \label{sec:NearlyIncomp}
Many polymers are much stiffer in their volumetric response than their deviatoric response. This necessitates methods which circumvent volumetric locking for cases close to the incompressibility condition. To this end, we decompose the strain energy density into volumetric and isochoric parts
\begin{equation} \label{FE_decomp}
	\tilde{W}(\boldsymbol{F})=\tilde{W}^{\text{iso}}(\boldsymbol{\bar{F}}) +W^{\text{vol}}(J) =W^{\text{iso}}(\boldsymbol{\bar{C}})  +W^{\text{vol}}(J) \, ,
\end{equation}
where $\boldsymbol{\bar{F}}= J^{-\frac{1}{3}}\boldsymbol{F}$ and $\boldsymbol{\bar{C}}= J^{-\frac{2}{3}}\boldsymbol{C}$ are isochoric (unimodular) tensors.
The deformation gradient can then be expressed as a multiplicative decomposition of volumetric and isochoric parts, i.e. $\boldsymbol{F}=\boldsymbol{\hat{F}}\boldsymbol{\bar{F}}$, where the volumetric part is given by $\boldsymbol{\hat{F}}=J^{\frac{1}{3}} \boldsymbol{\delta}$ \cite{flory1961thermodynamic}. 
A three field Hu-Washizu functional is used for the enforcement of nearly incompressible materials \cite{nagtegaal1974numerically}. A kinematic variable $\theta$ is introduced which represents dilatation, while a Lagrange multiplier $p$ is used to enforce $\theta=J$. Therefore, the internal potential energy of a single element is given by 
\begin{equation}
	\Pi_{\text{int}}^{\text{e}}(\boldsymbol{\phi},\theta,p) =\int_{\Omega_0^{\text{e}}} W^{\text{iso}}(\boldsymbol{\bar{C}}(\boldsymbol{\phi})) d \Omega_0 + \int_{\Omega_0^{\text{e}}} W^{\text{vol}}(\theta) d \Omega_0 + \int_{\Omega_0^{\text{e}}} p(J-\theta) d \Omega_0 \, .
\end{equation}
The stationary condition is found by setting the directional derivative of each argument to zero.
In the case of $p$ and $\theta$, we have
\begin{equation}
	\int_{\Omega_0^{\text{e}}} \big( J - \theta  \big) \delta p \, d \Omega_0 =0 \, ; \quad \quad
	\int_{\Omega_0^{\text{e}}} \big(\frac{ \partial W^{\text{vol}}(\theta)}{\partial\theta}-p  \big) \delta \theta \, d \Omega_0 =0 \, .
\end{equation}
One of the most commonly implemented elements is the $Q1$-$P0$ element \cite{wriggers2008nonlinear}, which employs a piecewise constant interpolation of both of these variables, leading to
\begin{equation}
	\theta =\frac{v_e}{V_e} = \frac{\int_{\Omega_t^{\text{e}}}d \Omega_t }{\int_{\Omega_0^{\text{e}}}d \Omega_0} ; \quad \quad \quad 	p=	\frac{ \partial W^{\text{vol}}(\theta)}{\partial\theta}.
\end{equation}
This is then used in conjunction with first-order Lagrange polynomials for the displacement on eight-noded isoparametric elements. Because they are local to the element, $p$ and $\theta$ can be statically condensed, leading to a generalized displacement method \cite{bonet1997nonlinear,simo1991quasi}.

\subsubsection{Volumetric Energy}
Polymeric materials tend to display less viscous behavior in their volumetric response versus their deviatoric response. Therefore a common assumption is that the volumetric part is purely (hyper) elastic \cite{holzapfel1996new,lion1997physically}, which is the approach taken here. The following simple form is used
\begin{equation}
	W^{\text{vol}}(\theta)= \frac{K}{2}\Big(\theta-1\Big)^2
\end{equation}
\begin{equation}
	\begin{split}
		p = \frac{\partial W}{\partial \theta}=K(\theta-1) \, ,
	\end{split}
\end{equation} 
where $K$ is a bulk modulus.

\subsubsection{Isochoric Yeoh} \label{sec:Yeoh}
For small to moderate deformation, the neo-Hookean model can be used to accurately model rubbers. However, under large tensile deformation, the stress strain relationship is often sigmoidal in shape. This behavior can be accommodated by utilizing a free energy that is cubic in $I_1$, as proposed by Yeoh \cite{yeoh1993some}. For the isochoric component, this leads to the following strain energy and deviatoric Cauchy stress
\begin{equation}
	\begin{split}
		W^{\text{iso}}= b_1 \big(\bar{I}_1-3\big) + b_2 \big(\bar{I}_1-3\big) ^2 + b_3 \big(\bar{I}_1-3\big)^3 = c_1 \bar{I}_1 + c_2 \bar{I}_1^2 + c_3 \bar{I}_1^3 +c_4 
	\end{split}
\end{equation} 
\begin{equation}
	\begin{split}
		\sigma^{\text{d}}_{ij}	=\frac{2c_1}{J}\Big( \bar{b}_{ij}  -\frac{1}{3} \bar{I}_1 \delta_{ij}\Big) +\frac{4c_2}{J}\Big( \bar{I}_1 \bar{b}_{ij}   -\frac{1}{3} \bar{I}^2_1 \delta_{ij}\Big)+\frac{6c_3}{J}\Big(\bar{I}_1^2 \bar{b}_{ij}   -\frac{1}{3}  \bar{I}_1^3 \delta_{ij}\Big) \, .
	\end{split}	
\end{equation}
The decomposition for the deviatoric part of the Yeoh Cauchy stress is found in Table \ref{Yeohtable}. Including only the first component leads to a different form of the neo-Hookean model, one that is most commonly implemented in finite element codes for hyperelasticity \cite{kossa2023analysis}. 
\begin{table}[h]
		\centering
	\begin{tabular}{ |r |l| } 
			\hline
				$A^{1}_{MN}  $ &$	J(s)\bar{C}_{MN}^{-1}(s)$ \\ 
			$A^{2}_{MNOP} $ & $ J(s)\bar{C}_{MN}^{-1}(s)\bar{C}_{OP}^{-1}(s)$\\ 
			$A^{3}_{MNOPQR}$&$	J(s)\bar{C}_{MN}^{-1}(s)\bar{C}_{OP}^{-1}(s)\bar{C}_{QR}^{-1}(s) $ \\
		\hline
			$	B^{1}_{MNij}    $ &$	\frac{2 c_1}{J(t)} \Big(  \bar{F}_{iM}(t)\bar{F}_{jN}(t) - \frac{1}{3}\delta_{ij}\bar{C}_{MN}(t) \Big)$    \\ 
		$ 	B^{2}_{MNOPij}    $ & $ 	\frac{4 c_2}{J(t)} \Big(  \bar{F}_{iM}(t)\bar{F}_{jN}(t)\bar{C}_{OP}(t) - \frac{1}{3}\delta_{ij}\bar{C}_{MN}(t)\bar{C}_{OP}(t) \Big)$ \\
		$ B^{3}_{MNOPQRij}  $&$\frac{6 c_3}{J(t)} \Big(  \bar{F}_{iM}(t)\bar{F}_{jN}(t)\bar{C}_{OP}(t)\bar{C}_{QR}(t) - \frac{1}{3}\delta_{ij}\bar{C}_{MN}(t)\bar{C}_{OP}(t)\bar{C}_{QR}(t) \Big) $\\
		\hline 
\end{tabular}
	\caption{Yeoh decomposition}
	\label{Yeohtable}
\end{table}

\section{Discussion and Numerical Examples} \label{sec:discuss}
\subsection{Multiple Network Generalization and Alternative Representation}
The first-order kinetics assumption is important for implementation into finite element methods. However, many materials are more adequately described by introducing multiple transient networks with differing detachment rates (i.e. a discrete spectrum of relaxation times). Generalizing to $M$ networks leads to
\begin{equation}
	\boldsymbol{\sigma} =  \sum_{a=1}^{M}  \rho   \gamma_a(t,0)\frac{\partial \psi_a(\boldsymbol{F})}{\partial \boldsymbol{F}}\boldsymbol{F}^T +  \sum_{a=1}^{M}  \rho \intu_{0}^{t}  \gamma_a(t,s)\frac{\partial \psi_a(\boldsymbol{F}(t,s))}{\partial \boldsymbol{F}(t,s)}\boldsymbol{F}^T(t,s) ds \, ,
\end{equation}
where each $\gamma_a$ are governed by their own characteristic kinetics. Note that $\psi_a$ may be specified independently to impart different initial chain concentrations $\xi_0$ using the same functional form for $\psi_C$, or to use entirely different functional forms altogether. 

If any one of the characteristic rates are set to zero (or low compared to the time scales of interest), that network causes the material to behave like a viscoelastic solid, as it acts to bring the material back to the original reference configuration when unloaded. Otherwise, the material behaves like a viscoelastic fluid, allowing residual strain to develop over time. 
In fact, an analogous formulation (for a single network) can be derived via
\begin{equation} \label{AltFormFreeEnergy}
	\psi(t)=\intu_{-\infty}^{t}\gamma(t,s)\psi(\boldsymbol{F}(t,s))ds \, ,
\end{equation}
leading to the following expression for stress
\begin{equation} \label{eqn:altstress}
	\boldsymbol{\sigma} =  \rho \intu_{-\infty}^{t}  \gamma(t,s)\frac{\partial \psi(\boldsymbol{F}(t,s))}{\partial \boldsymbol{F}(t,s)}\boldsymbol{F}^T(t,s) ds  \, .
\end{equation}
This representation is similar to various other viscoelastic fluid models in the literature, and is equivalent to the transient network definition under the assumption that before the initial reference time at $t=0$, the material is undeformed ($\boldsymbol{F}(0,s)=\boldsymbol{I} \quad s<0$). 
In particular, the K-BKZ model  \cite{bernstein1963study,tanner1988bk} consists of hereditary integrals with limits approaching $-\infty$ whose kernels contain derivatives of potentials with respect to relative deformation gradients. An equivalence to transient network theory can be made if the ``memory functions'' of the K-BKZ model can be tied to the evolution of chain densities. Common implementation techniques for the fluid models include approximating the integral with a finite number of past deformations or establishing recurrence relations using approximations of the kernel through Taylor series expansions in time \cite{feng1992recurrence,erchiqui2005thermodynamic,keunings2003finite,Marin2009}.

\subsection{Comparison to Internal Variable Approaches}
A distinct but related internal variable approach was proposed by Lubliner \cite{Lubliner1985}, which is based on decomposing the deformation gradient into elastic and inelastic parts through
\begin{equation} \label{Lubliner1}
	\boldsymbol{F} =\boldsymbol{F}_e \boldsymbol{F}_i  ; \quad \boldsymbol{\bar{F}} = \boldsymbol{\bar{F}}_e \boldsymbol{F}_i \, .
\end{equation}
Notably, the inelastic part of the deformation gradient is presumed to remain isochoric (i.e. $\det \boldsymbol{F}_i=\det \boldsymbol{C}_i=1$).
The following additive split of the free energy is performed (similar to Equation \ref{FE_decomp}) that includes an internal variable $\boldsymbol{\zeta}$
\begin{equation} \label{Lubliner2}
	\tilde{W}(\boldsymbol{F},\boldsymbol{\zeta}) = \tilde{W}^{\text{iso}}(	\boldsymbol{\bar{F}},\boldsymbol{\zeta}) +W^{\text{vol}}(J) \, ,
\end{equation}
where isothermal conditions are assumed. The internal variable is chosen to be $\boldsymbol{\zeta}=\boldsymbol{C}_i^{-1}$. Then, by noting that invariants of $\boldsymbol{\bar{C}}_e$ are the same as invariants of $\boldsymbol{\bar{C}}\boldsymbol{C}_i^{-1}$, Lubliner introduced the following form for the free energy
\begin{equation} \label{Lubliner3}
	\tilde{W}(\boldsymbol{F},\boldsymbol{\zeta}) = W^{\text{iso}}(\boldsymbol{\bar{C}}\boldsymbol{C}_i^{-1}) +W^{\text{vol}}(J) \, .
\end{equation}
$W^{\text{iso}}$ and $W^{\text{vol}}$ are left general, allowing for a variety of functional forms to be employed. For the particular case of the isochoric neo-Hookean model, this leads to the following free energy and Cauchy stress
\begin{equation} \label{Lubliner4}
	W^{\text{iso}}(\boldsymbol{\bar{C}}\boldsymbol{C} _i^{-1}) = \frac{\mu}{2} \Big(  \Tr\big(\boldsymbol{\bar{C}}\boldsymbol{C}_i^{-1}\big) -3 \Big)
\end{equation}
%
%

\begin{equation}
	\boldsymbol{\sigma}^{\text{d}} = \frac{\mu}{J} \Big( \boldsymbol{\bar{F}} \boldsymbol{C}_i^{-1} \boldsymbol{\bar{F}}^T - \frac{1}{3} \left( \boldsymbol{C}_i^{-1} : \boldsymbol{\bar{C}} \right) \boldsymbol{\delta} \Big) \, .
\end{equation}
By comparing with Table \ref{Yeohtable}, we see that $\boldsymbol{C}_i^{-1}(t)$ can be reinterpreted as the history variable $\boldsymbol{\mathcal{H}}^1$ from Section \ref{sec:Yeoh} if we utilize the following evolution equation for $\boldsymbol{\bar{C}}_i^{-1}(t)$ (consistent with Equation \ref{rateformf})
\begin{equation}
	\begin{split}
		\dot{\boldsymbol{C}}_i^{-1}  = k \Big( J\boldsymbol{\bar{C}}^{-1}-\boldsymbol{C}_i^{-1}\Big) \, ,
	\end{split}
\end{equation} 
where
\begin{equation} \label{Lubhist}
	\begin{split}
		\boldsymbol{C}_i^{-1}(t) = \intu_{0}^{t} \gamma(t,s)J(s)\boldsymbol{\bar{C}}^{-1}(s) ds = \boldsymbol{\mathcal{H}}^1\, .
	\end{split}
\end{equation}
Lubliner's model differs slightly from the present approach, in that his rate equation for the internal variable is given by 
\begin{equation}
	\begin{split}
		\dot{\boldsymbol{C}}_i^{-1} = k \Big( \boldsymbol{\bar{C}}^{-1}-\boldsymbol{C}_i^{-1}\Big) \, .
	\end{split}
\end{equation}
Because Lubliner assumed that the inelastic component is isochoric, he suggests that his evolution equation may be viewed as a linearization of some nonlinear evolution equation that can enforce $\det \boldsymbol{F}_i=1$. Indeed, subsequent works based on the multiplicative split into elastic and inelastic part employ nonlinear evolution equations that respect $\det \boldsymbol{F}_i=1$ (e.g. \cite{bonet2001large, lefevre2024abaqus, wijaya2023unified}). However, no such interpretation or constraint is required for $\boldsymbol{\mathcal{H}}^1$ in the transient network approach. 
Rather, the contraction of $\boldsymbol{\mathcal{H}}^1:\boldsymbol{B}^1$ represents the (neo-Hookean) stress emanating from the isochoric parts of the deformation gradients relative to a distribution of past deformation states.

While there is a clear benefit to the simplicity of the evolution of the history variables, the kernel decomposition needed to apply this approach is fairly restrictive. For the most part, invariant-based mechanical free energy definitions that allow for this decomposition involve polynomials. This is perhaps more restrictive than the need to use first-order kinetics, which can be overcome in some sense by using a discrete spectrum of characteristic rates. Additionally, the higher the order of the polynomial, the higher rank tensors are needed to store these history variables. For example, the Yeoh model requires 2, 4, and 6-rank tensors as history variables, corresponding to first, second, and third order polynomial terms. While this can be prohibitive in terms of both computing time and memory requirements, these tensors have both major and minor symmetries that can be exploited. Indeed, the total number of components for these three particular tensors are 9, 81, and 729, reducing to 6, 21, and 56 unique values, respectively. 

The stretch-based transient networks defined here do not have the same problem, as the storage and computational requirements remain independent of the order of the free energy terms. Therefore, for highly nonlinear stress strain responses, the stretch-based models are an attractive alternative. Though not pursued here, one may also consider functional forms that involve relative invariants (i.e. $\frac{I_1(t)}{I_1(s)},\frac{I_2(t)}{I_2(s)},\frac{I_3(t)}{I_3(s)}$), rather than invariants of the relative deformation gradient. This would mitigate the storage and computation issues as well.  
    

\subsection{Representative Examples} \label{section:subexamples}
In this section, multiple examples are presented to illustrate the ability to model stress relaxation and age-induced permanent set. 
All of the problems are solved quasi-statically, meaning the only time dependence is through the kinetics associated with the transient networks, as well as any loading conditions. The material models outlined in the previous section will be used throughout, possibly with multiple networks. The neo-Hookean, Blatz-Ko and Ogden-Hill parameters are chosen to represent highly compressible foams, while the Yeoh parameters are chosen to represent nearly incompressible polymers. 
\subsubsection{Uniaxial Response}
\begin{figure} [h]
	\centering
   \begin{subfigure}[t]{0.48\textwidth}
	\includegraphics[width =1.0\textwidth]{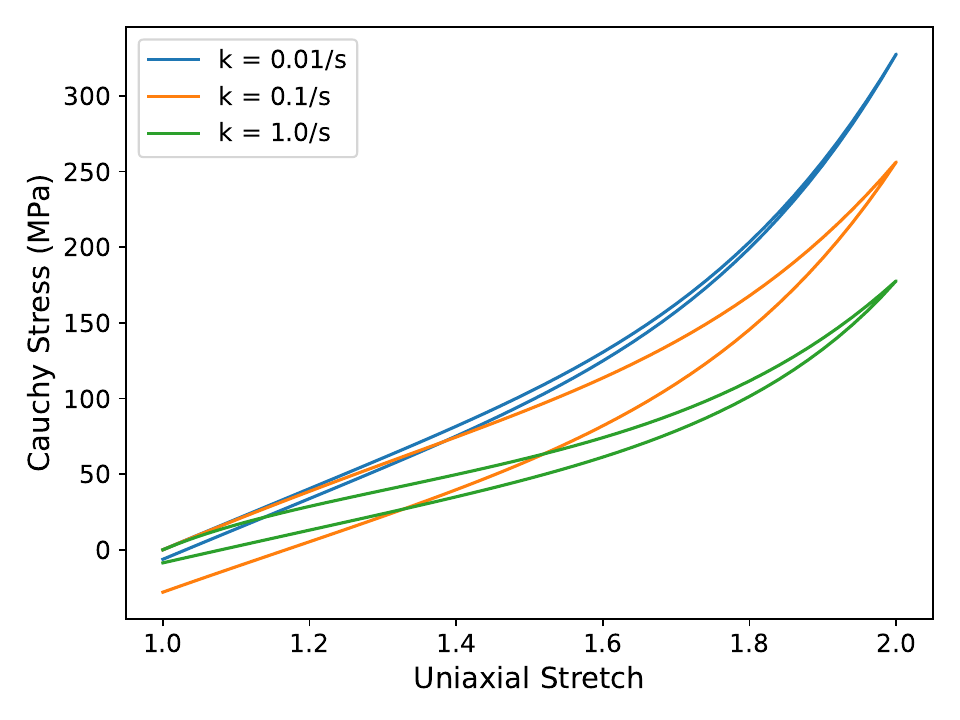}
	\caption{Cyclic Loading of a Yeoh material with one permanent and one transient network: Comparison of uniaxial stress versus stretch for three separate choices of the transient network's characteristic rate.}
		\label{fig:Yeoh1}
	\end{subfigure}\hspace{5mm}
  \begin{subfigure}[t]{0.48\textwidth}
	\includegraphics[width=1.0\textwidth]{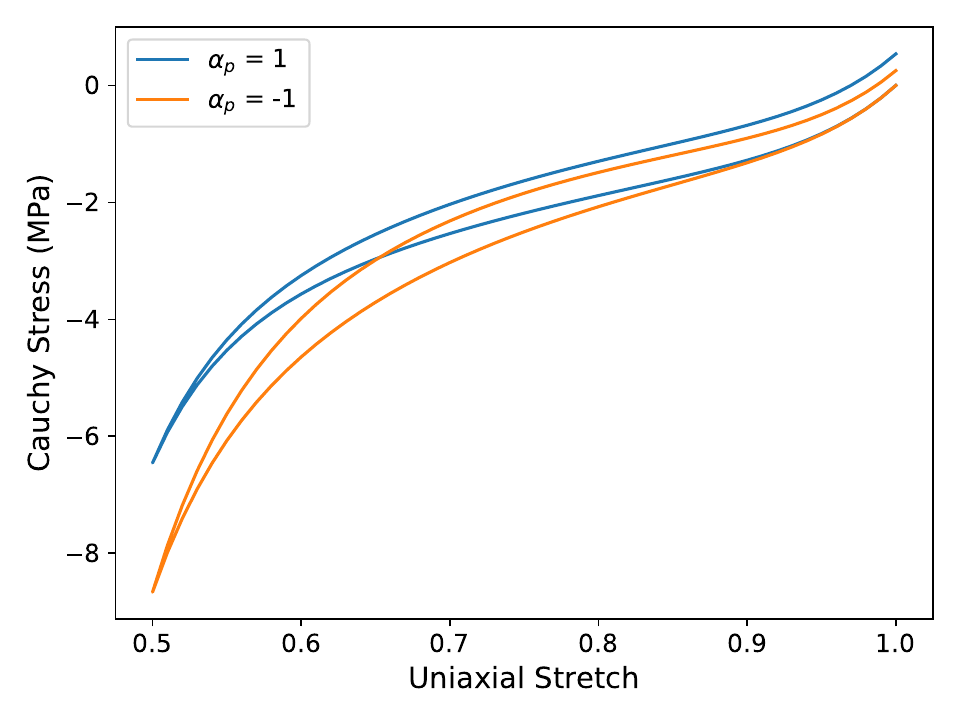}
	\caption{Cyclic Loading of an Ogden-Hill foam with one permanent and one transient network: Comparison of uniaxial stress versus stretch for different choices of the transient network's strain measure.}
		\label{fig:Ogden1}
	\end{subfigure}
\end{figure}

This pair of examples demonstrates the viscoelastic response of uniaxial compression and tension tests performed on a cube with $10$ mm sides. 
In the first example, the cube is pulled in one direction to double its original length through displacement boundary conditions, and then returned to its original length, with the loading cycle taking $20s$. The material is represented by the nearly incompressible Yeoh model, with coefficients $(c_1, c_2, c_3)=(50, -10, 1)$ MPa and a bulk modulus $K = 10,000$ MPa. The material has two networks with these parameters, one that is permanent (i.e. $k=0$) and one with a specified relaxation rate. There are two limiting cases for the second network based on the choice of $k$. When $k \rightarrow 0$, there is no degradation, so that it acts as a permanent network, and the effective stiffness is twice that of the first network. Conversely, when $k \rightarrow \infty$, the degradation is so fast that its stiffness contribution is negligible. For intermediate values of these limiting cases, a comparison of the measured Cauchy stress as a function of strain is shown in Figure \ref{fig:Yeoh1}. 

Next, a highly compressible cube of the same dimensions is compressed in one direction to half its original length and and then back to its original configuration, with the loading cycle taking $20s$. The material is taken to have one permanent Ogden-Hill network with parameters $(\alpha_1, \alpha_2)=(-5, 10)$ and $(\mu_1, \mu_2)=(-0.001, 1.0)$ MPa, chosen to represent the softening due to elastic buckling of foams as well as the subsequent stiffening due to densification. A transient network with $\alpha_3=\pm 1.0$ and $\mu_3= \pm 1.0$ MPa is also utilized. Figure \ref{fig:Ogden1} showcases the different behavior that can be obtained based on the two different choices of the transient network's strain measure.
  

\subsubsection{Age-induced Permanent Set of O-rings Under Static Loading}
\begin{figure} [!htb]
	\centering
	\begin{subfigure}[t]{0.15\textwidth}
		\includegraphics[trim={0cm 1cm 16.7cm 0cm},clip,width=\textwidth]{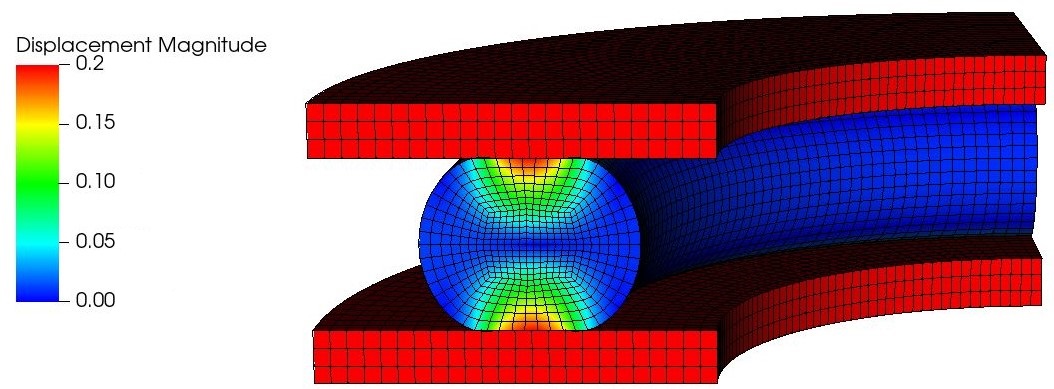}
		\nonumber
	\end{subfigure}
    \begin{subfigure}[t]{0.34\textwidth}
		\includegraphics[trim={6cm 0cm -1cm 0cm},clip,width=\textwidth]{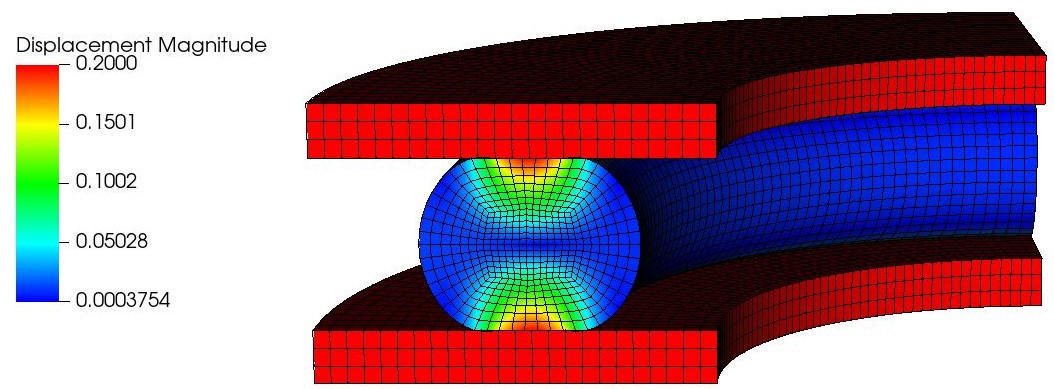}
		\caption{Blatz-Ko}
	\end{subfigure}
	\begin{subfigure}[t]{0.15\textwidth}
	\includegraphics[trim={0cm 1.8cm 16.7cm 0cm},clip,width=\textwidth]{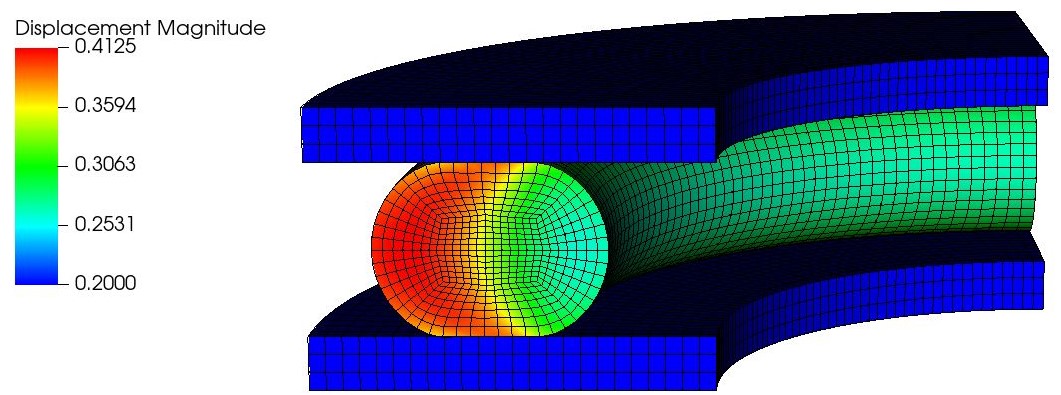}
	\nonumber
	\end{subfigure}
	\begin{subfigure}[t]{0.34\textwidth}
		\includegraphics[trim={6cm 0cm -1cm 0cm},clip,width=\textwidth]{CompressedYeoh}
	    \caption{Yeoh}
	\end{subfigure}
	\caption{Displacement magnitude ($mm$) for fully compressed state}
	\label{fig:BKYComp}
\end{figure}

\begin{figure} [!htb]
	\centering
	\begin{subfigure}[t]{0.1\textwidth}
		\includegraphics[trim={0cm 1.6cm 17cm 0.1cm},clip,width=\textwidth]{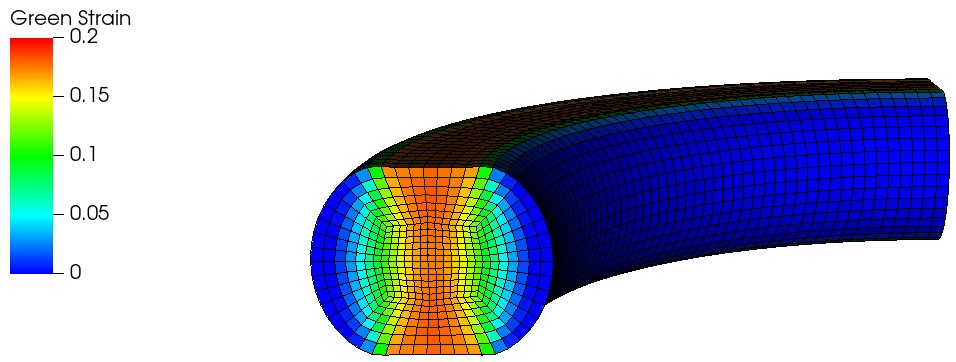}
		\nonumber
	\end{subfigure}
	\begin{subfigure}[t]{0.29\textwidth}
		\includegraphics[trim={6cm 0cm 0cm 1cm},clip,width=\textwidth]{BK_GS_compressed}
		\caption{Fully compressed state}
	\end{subfigure}
	\begin{subfigure}[t]{0.29\textwidth}
	\includegraphics[trim={6cm 0.0cm 0cm 1cm},clip,width=\textwidth]{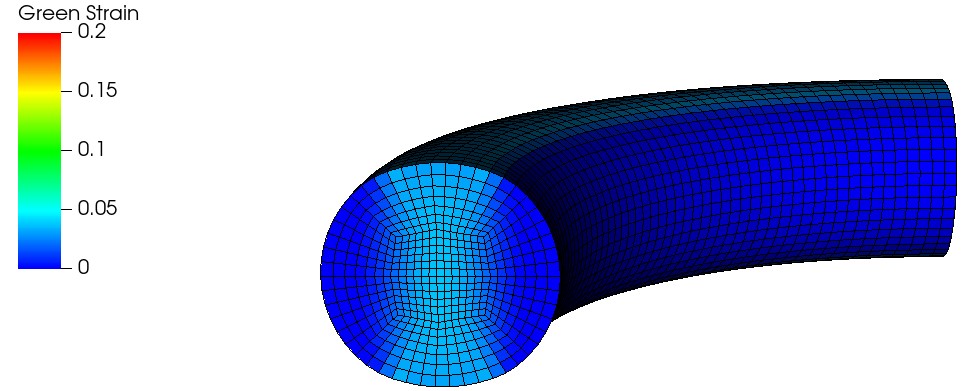}
	\caption{Residual strain: 2 years}
	\end{subfigure}
	\begin{subfigure}[t]{0.3\textwidth}
	\includegraphics[trim={6cm 0cm 0cm 1cm},clip,width=\textwidth]{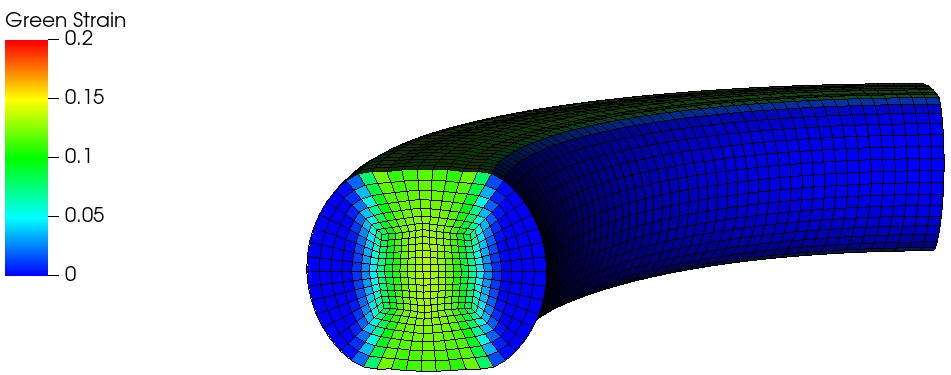}
	\caption{Residual strain: 10 years}
\end{subfigure}
	\caption{Aging of Blatz-Ko foam: Green strain $E_{zz} =\frac{1}{2}(C_{zz}-1.0)$ in loaded and unloaded states}
	\label{fig:BKAge1}
\end{figure}

\begin{figure} [!htb]
	\centering
	\begin{subfigure}[t]{0.1\textwidth}
	\includegraphics[trim={0cm 0.0cm 0.0cm 0.0cm},clip,width=\textwidth]{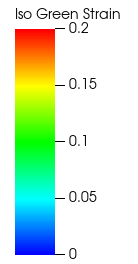}
	\nonumber
    \end{subfigure}
	\begin{subfigure}[t]{0.29\textwidth}
		\includegraphics[trim={5.8cm 0cm 0cm 1cm},clip,width=\textwidth]{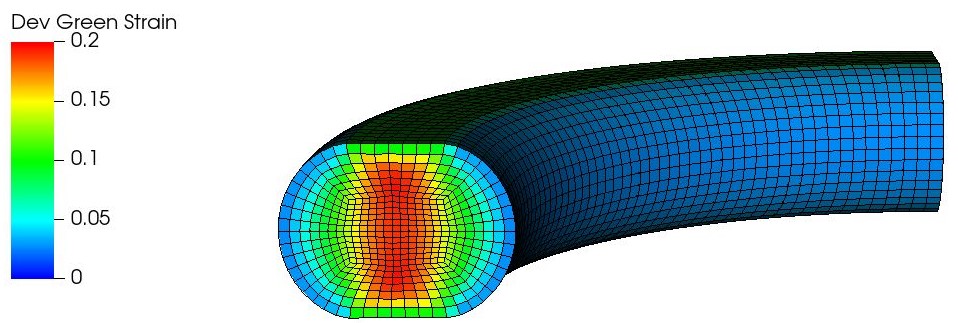}
		\caption{Fully compressed state}
	\end{subfigure}
	\begin{subfigure}[t]{0.29\textwidth}
		\includegraphics[trim={6cm 0.02cm 0cm 1cm},clip,width=\textwidth]{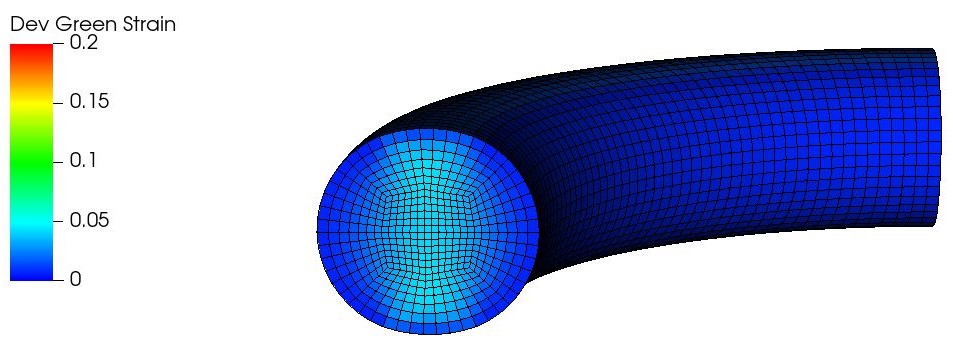}
	\caption{Residual strain: 2 years}
	\end{subfigure}
	\begin{subfigure}[t]{0.29\textwidth}
		\includegraphics[trim={6cm 0cm 0cm 1cm},clip,width=\textwidth]{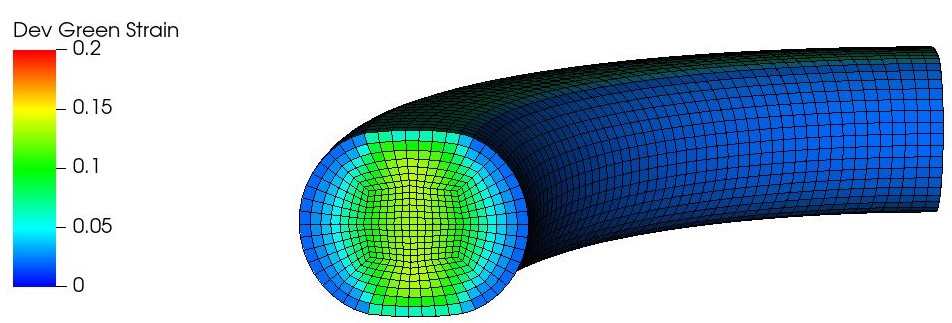}
	\caption{Residual strain: 10 years}
	\end{subfigure}
	\caption{Aging of Yeoh polymer: Isochoric Green strain $\bar{E}_{zz}=\frac{1}{2}(\bar{C}_{zz}-1.0)$ in loaded and unloaded states}
	\label{fig:YAge1}
\end{figure}

The next example is meant to demonstrate age-induced residual deformation due to set strains in a three dimensional setting. A polymeric o-ring is compressed between two rigid plates for an extended period of time before being removed.  Two different materials are presented for comparison, a Blatz-Ko material meant to represent a foam and a Yeoh material to represent a solid polymer. For both cases, a single transient network is used, with a constant decay rate of $k=0.1\frac{1}{\text{yrs}}$ , allowing for residual strain to occur. The parameters for the Yeoh network are the same as the previous example, while the Blatz Ko network has parameters $f=0.5$, $\mu=2.0$ MPa, and $\beta=0.05$. The o-ring torus has major radius of 5 mm and minor radius of 1 mm. Quarter symmetry boundary conditions are imposed, with the relative displacement between the two rigid plates set to 0.4 mm, as shown in Figure \ref{fig:BKYComp}. The difference in compressibility is especially apparent when comparing the deformation in this compressed state, where the Yeoh material has much more radial expansion as compared to the Blatz-Ko foam. The effect of set strains for these materials after two and ten years is shown in Figures \ref{fig:BKAge1} and \ref{fig:YAge1}. 


\subsubsection{Temperature and History Dependent Response}
The aim of this final pair of examples is to explicitly show the ability to model both short term viscoelasticity and age-induced permanent set, as well as demonstrate temperature dependence on the kinetics. For both problems, a rigid half-cylinder acts as an indenter into a compressible neo-Hookean material with parameters $\lambda=0.03$ MPa and $\mu=1.5$ MPa. A temperature gradient is imposed in which the block's left side is held at 273K and its right side at 373K, with a linear interpolation in between. Temperature dependent kinetics are then incorporated via the following Arrhenius equation
\begin{equation}
k=A\exp\Big(\frac{-E_A}{RT}\Big) \, ,
\end{equation}
where $R=8.314\frac{\text{J}}{ \text{K mol}}$ is the gas constant, $A$ is the pre-exponential factor and $E_A$ is the activation energy. Two networks are specified, both having an activation energy of $E_A=10\frac{\text{kJ}}{ \text{ mol}}$. However, the networks are very disparate in their characteristic times, with pre-factors of $A=20\frac{1}{\text{s}}$ and $A=20\frac{1}{\text{yrs}}$, respectively. 

\begin{figure}[h]
	\centering
	\begin{subfigure}[]{0.90\textwidth}
		\centering
		\includegraphics[trim={5cm 50cm 78cm 10cm},clip,width=0.2\textwidth]{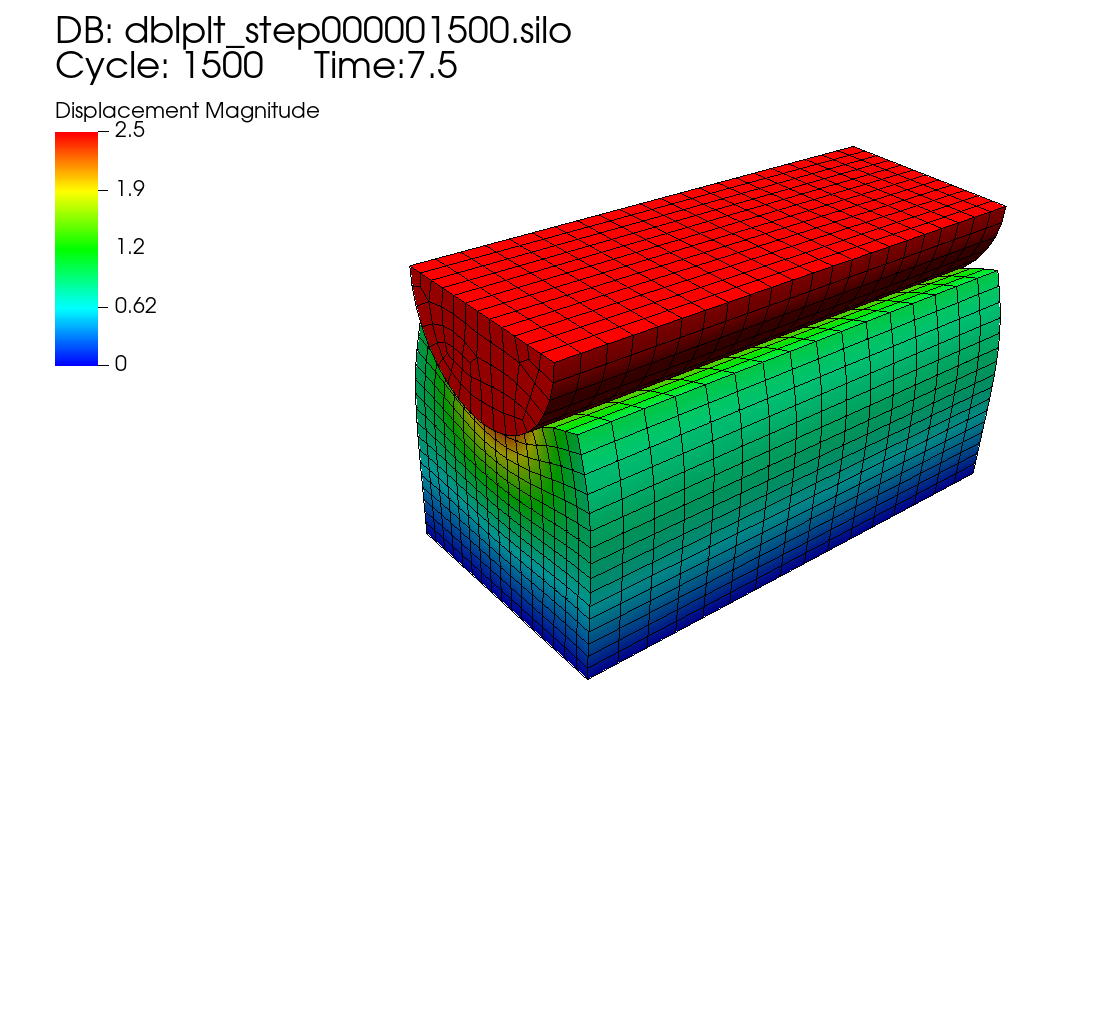}
        \includegraphics[trim={39cm 27cm 8cm 12cm},clip,width=0.3\textwidth]{NEWCompState0000}
	\caption{Deformation state at full indentation of neo-Hookean material}
	\label{fig:IndentState}
\end{subfigure}
	\begin{subfigure}[t]{0.990\textwidth}
  		\centering
		\includegraphics[trim={5cm 60cm 78cm 10cm},clip,width=0.17\textwidth]{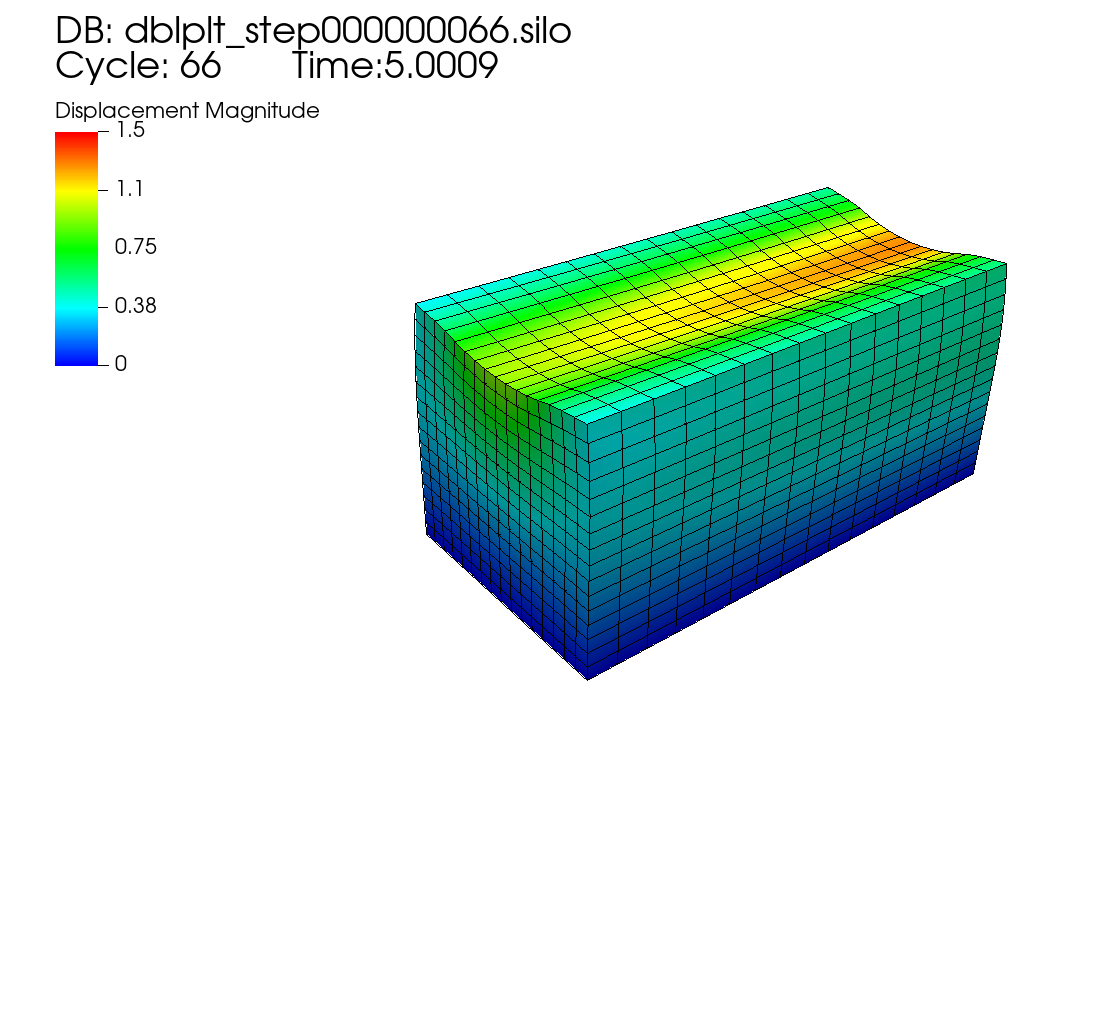}
		\includegraphics[trim={39cm 27cm 8cm 12cm},clip,width=0.20\textwidth]{NEWRelax0000}
		\includegraphics[trim={39cm 27cm 8cm 12cm},clip,width=0.20\textwidth]{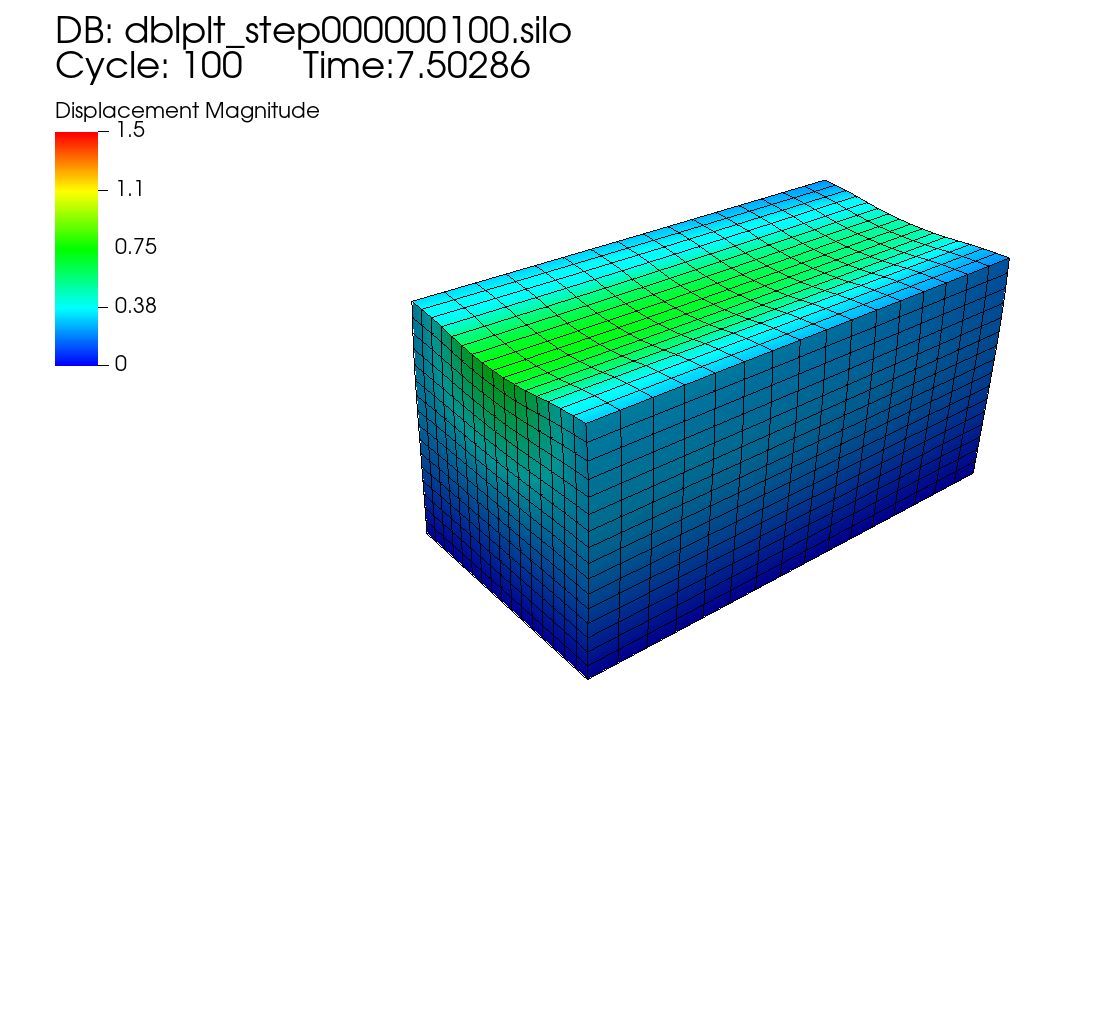}
		\includegraphics[trim={39cm 27cm 8cm 12cm},clip,width=0.20\textwidth]{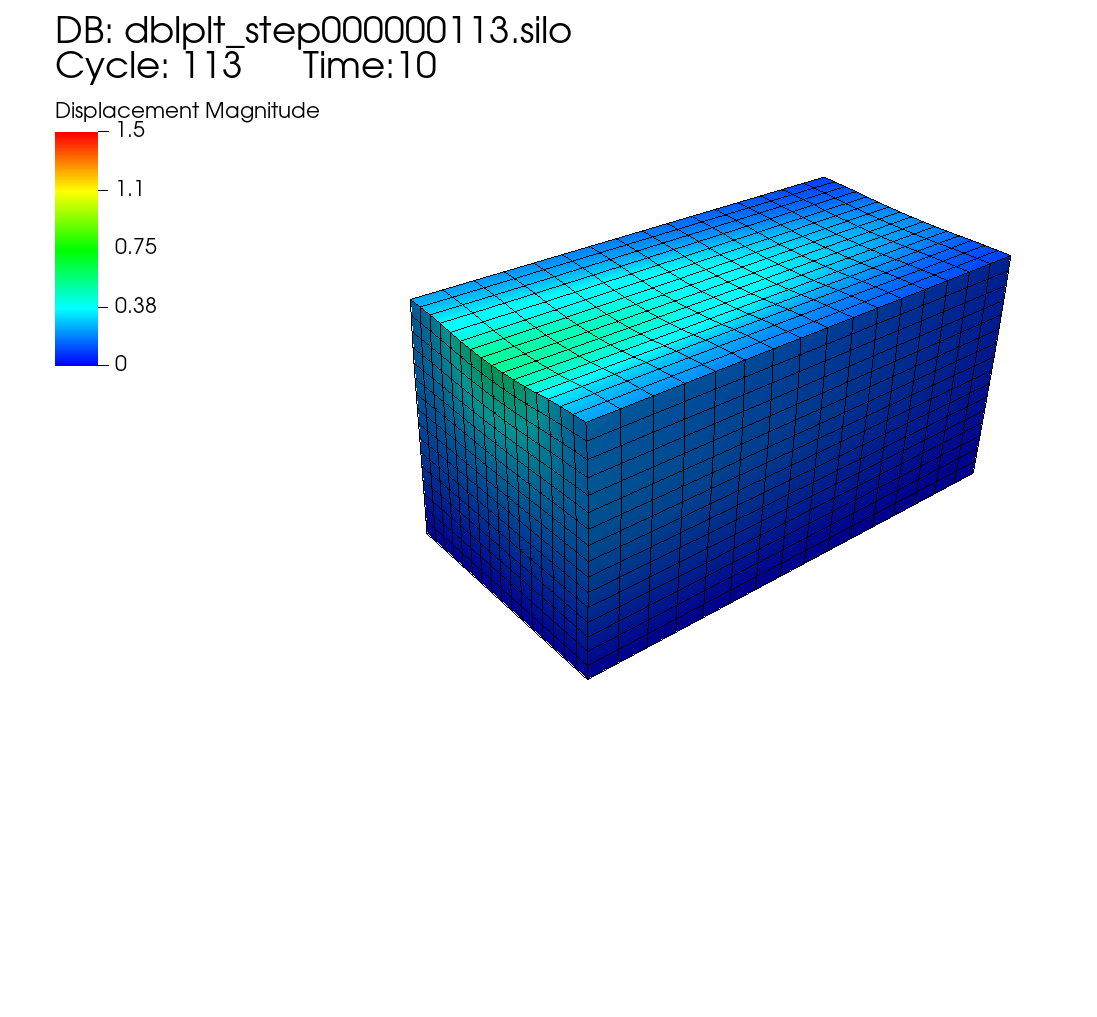}
		\includegraphics[trim={39cm 27cm 8cm 12cm},clip,width=0.20\textwidth]{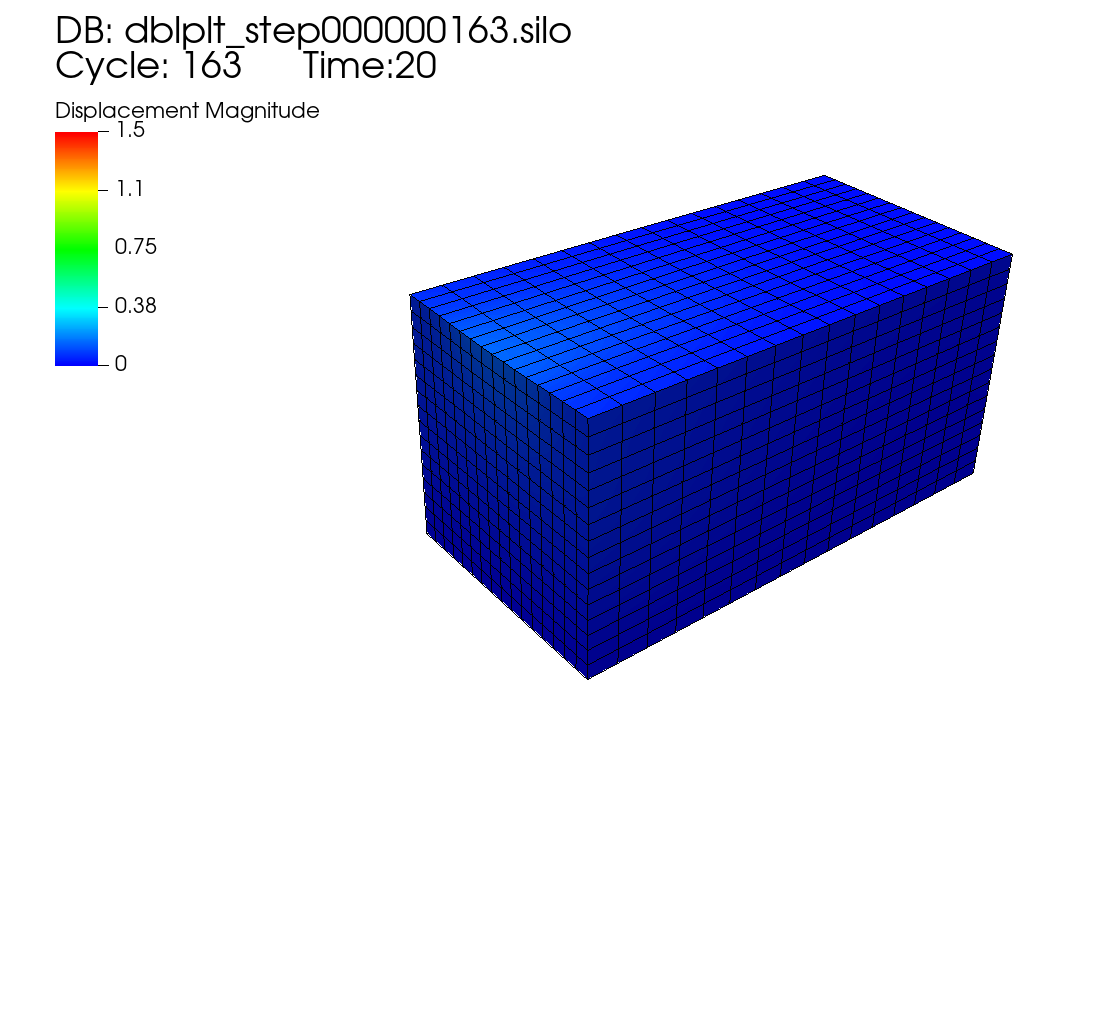}
		\caption{Viscoelastic response: 0s, 2.5s, 5s, and 15s after release}
		\label{fig:NHrelax1}
	\end{subfigure}
	\begin{subfigure}[t]{0.99\textwidth}
		\centering
		\includegraphics[trim={5cm 60cm 78cm 10cm},clip,width=0.17\textwidth]{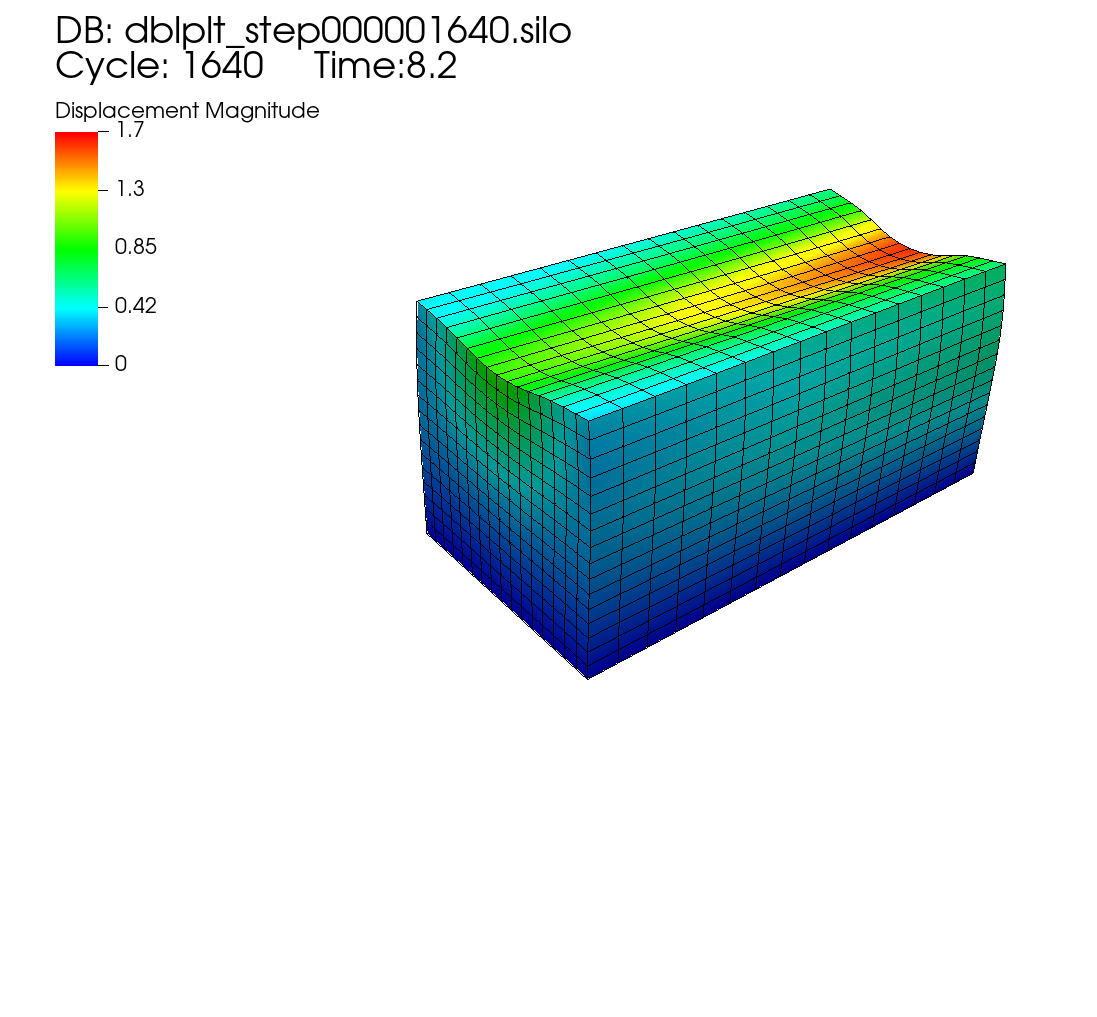}
		\includegraphics[trim={39cm 27cm 8cm 12cm},clip,width=0.20\textwidth]{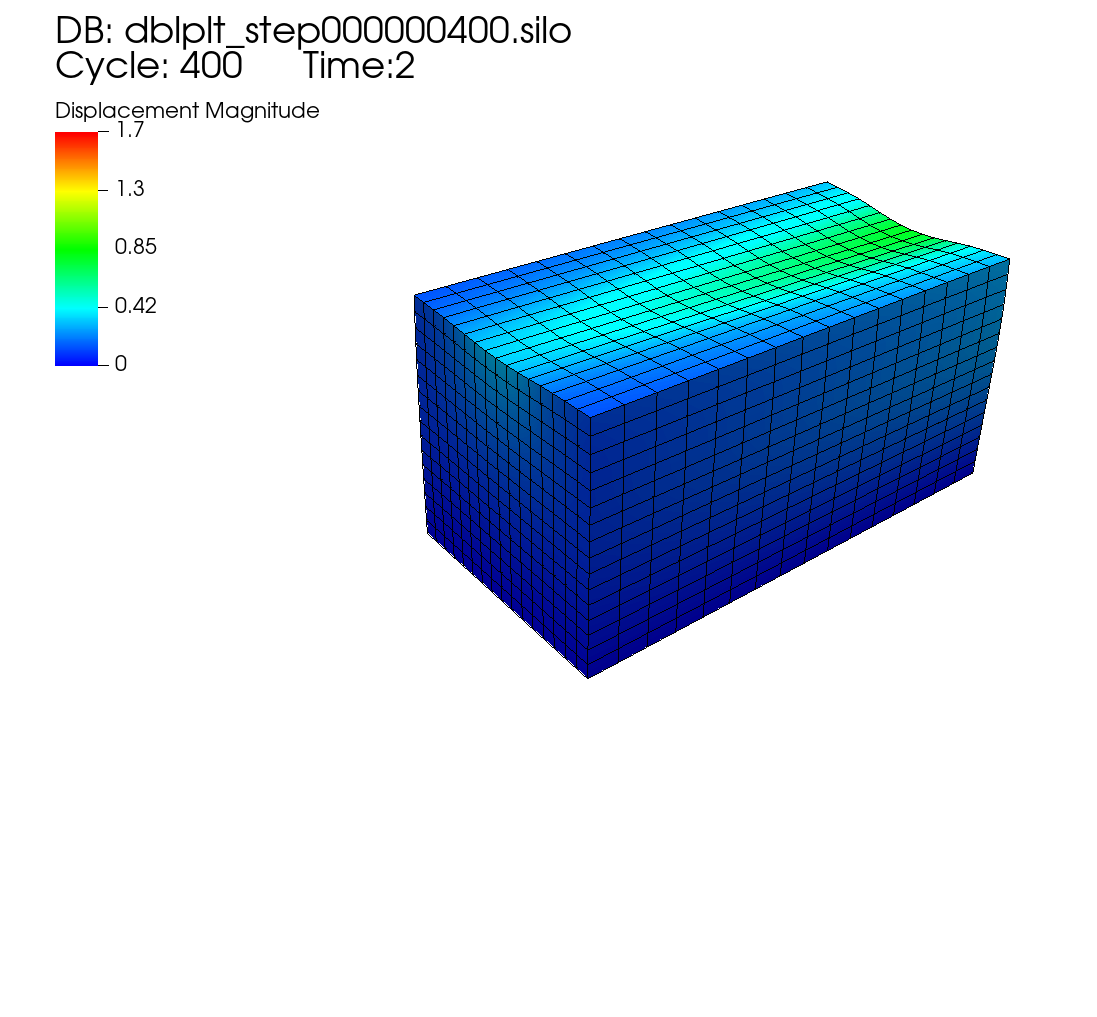}
		\includegraphics[trim={39cm 27cm 8cm 12cm},clip,width=0.20\textwidth]{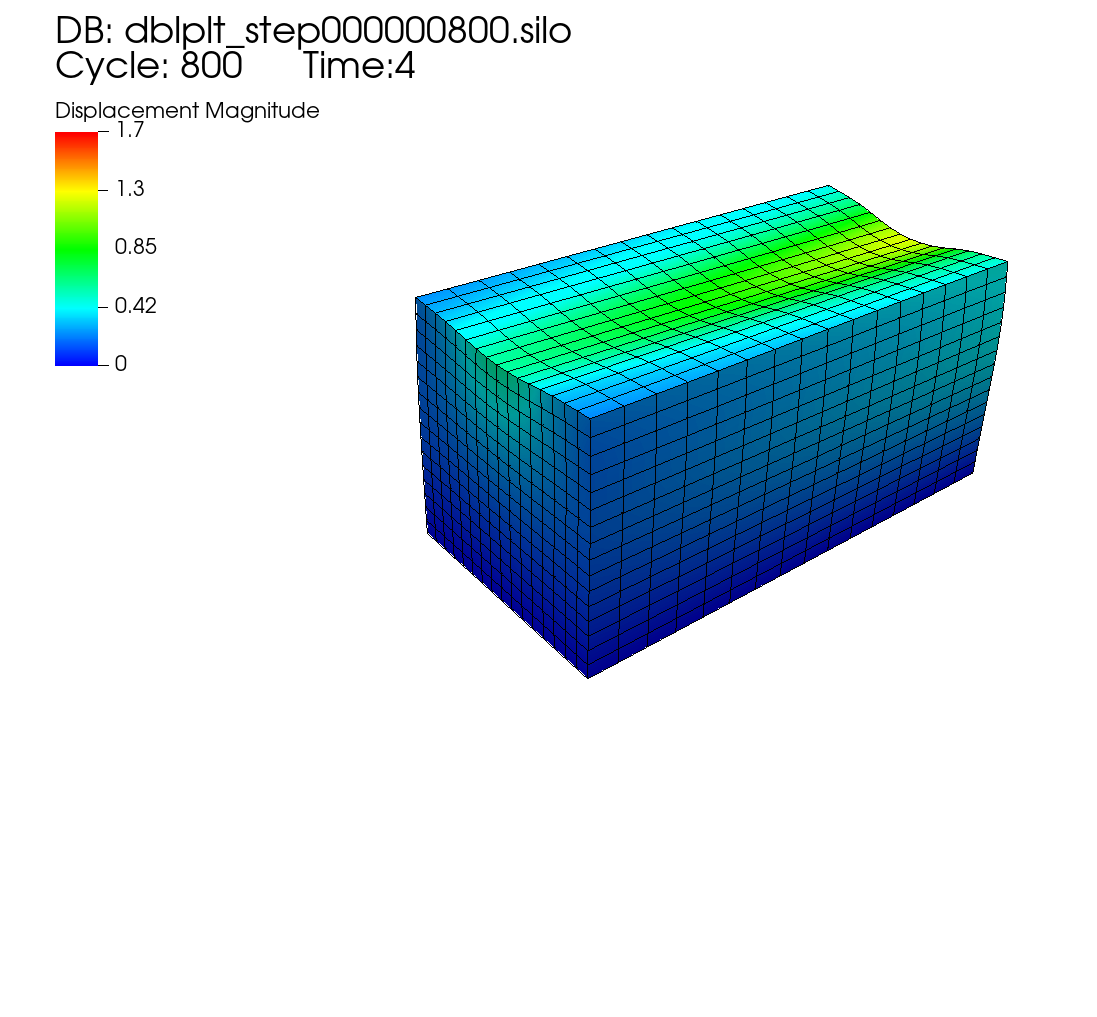}
		\includegraphics[trim={39cm 27cm 8cm 12cm},clip,width=0.20\textwidth]{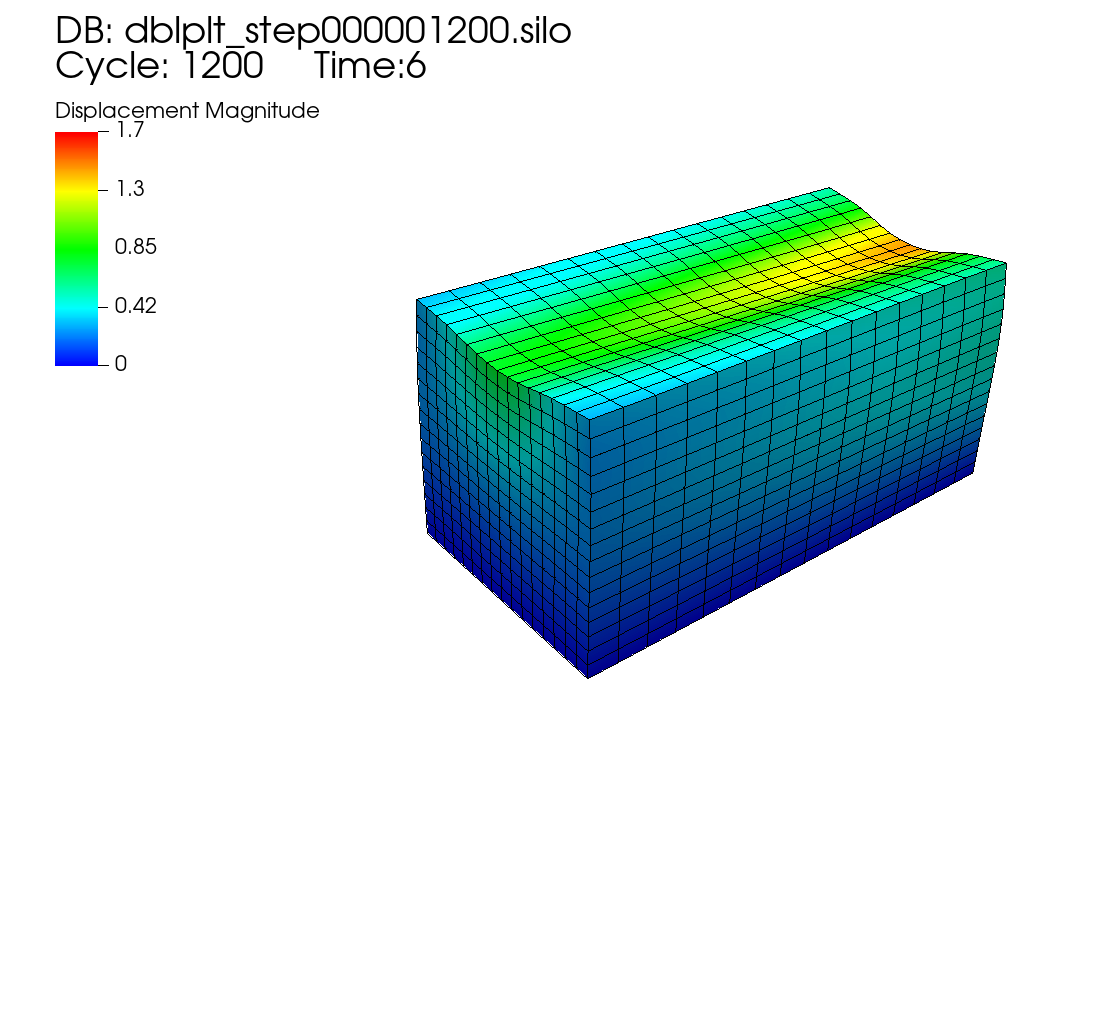}
		\includegraphics[trim={39cm 27cm 8cm 12cm},clip,width=0.20\textwidth]{NEWPermSet0000}		
     	\caption{Cyclic indentation: Residual deformation after 5, 10, 15, and 20 cycles, corresponding to 2, 4, 6, and 8 years}
   		\label{fig:NHcycle1}
	\end{subfigure}
\caption{Displacement magnitudes (mm) for neo-Hookean indentation examples}
\end{figure}

In the first example, the cylinder is rapidly indented into the compressible block and held there for a total of $5s$, then immediately removed (loading and unloading times take $0.001s$). 
Upon unloading, the block remains fairly compressed, then relaxes back to its original configuration over time, as seen in Figure \ref{fig:NHrelax1}. Effectively, the first network stays intact, as the loading conditions are orders of magnitude faster than its degradation rate. The hotter parts of the material start with a larger deformation immediately after unloading, as a larger fraction of the second network's equilibrium state has become associated with the fully compressed state shown in Figure \ref{fig:IndentState}. However, the faster kinetics associated with higher temperatures also causes these parts of the block to return to their original configuration faster. 

In the second example, the rigid cylinder is repeatedly indented to the configuration in Figure \ref{fig:IndentState} and removed through displacement boundary conditions that linearly ramp up and down with time. Each full loading and unloading cycle is set to a period of $0.4$ years. In this case, the second network's kinetics are so fast relative to the loading condition that its effect is negligible, and the residual deformation stems from degradation of the first network, as seen in Figure \ref{fig:NHcycle1}. As one may expect, it is apparent that the aging effects are more pronounced where the temperature is highest.


\section{Summary}
A hereditary integral framework was presented using transient network theory. This is a continuous extension of the discrete two-network Tobolsky model. In this theory, the material relaxes due to polymer chains detaching from associated networks. Meanwhile new networks are created in a stress free state, and the free energy of the new network depends on the relative deformation gradient. While a number of models have been developed based on this premise, their adoption into finite element codes has been hindered by the computational expense of storing and evaluating the hereditary integral. In this paper, we demonstrate that under the assumption of first-order degradation kinetics, a decomposition of the kernel for various free energies allows for a recurrence relationship to be established, bypassing the need to integrate over all time history. This greatly simplifies the numerical implementation. 

Practically, materials can be modeled as a sum of multiple transient networks with different kinetics. If any of the networks are made permanent, 
the material acts as a viscoelastic solid, and no true permanent deformation occurs. This is often a desirable feature when modeling short term viscoelastic behavior.
Alternatively, the use of nonzero degradation kinetics for all networks allows for residual strain to occur, which can be useful for modeling permanent set. This is especially true with regards to thermal-aging applications, for which the Tobolsky model remains a popular approach to this day.

\section*{Acknowledgement}
This work was performed under the auspices of the U.S. Department of Energy by Lawrence Livermore National Laboratory under the Contract DE-AC52-07NA27344, Release number LLNL-JRNL-871273. J.B.F. was supported by the U.S. Department of Energy, Office of Science, Office of Advanced Scientific Computing Research, Department of Energy Computational Science Graduate Fellowship under Award DE-SC0023112.

\appendix

\section{Relative Deformation Gradient Identities} \label{section:RelIdentities}
The relative deformation gradient can be expressed in terms of the standard deformation gradients at time $s$ and $t$ through $\boldsymbol{F}(t,s)=\boldsymbol{F}(t)\boldsymbol{F}^{-1}(s)$. For indicial notation, components associated with the configuration at times $0$, $t$ and $s$ are denoted by uppercase, lowercase, and calligraphic letters, respectively. For decomposing the transient network stresses and tangents, the following identities may be useful:\\
\newline
Right Green Cauchy Tensor:
\begin{equation}
	C_{\mathcal{I}\mathcal{J}}(t,s)=F_{kM}(t)F_{M\mathcal{I}}^{-1}(s)F_{kN}(t)F_{N\mathcal{J}}^{-1}(s)=C_{MN}(t)F^{-1}_{M\mathcal{I}}(s)F^{-1}_{N\mathcal{J}}(s)
\end{equation}
Left Cauchy-Green Tensor:
\begin{equation}
	b_{ij}(t,s)=F_{iM}(t)F_{M\mathcal{K}}^{-1}(s)F_{jN}(t)F_{N\mathcal{K}}^{-1}(s)=F_{iM}(t)F_{jN}(t)C_{MN}^{-1}(s)
\end{equation}
First Invariant
\begin{equation}
	I_1(t,s)=C_{\mathcal{I}\mathcal{I}}(t,s)=C_{MN}(t)C^{-1}_{MN}(s)
\end{equation}
Second Invariant
\begin{equation}
	\begin{split}
		I_2(t,s)=&\frac{1}{2}\Big(C_{\mathcal{I}\mathcal{J}}(t,s)C_{\mathcal{I}\mathcal{J}}(t,s)-C_{\mathcal{I}\mathcal{J}}(t,s)C_{\mathcal{J}\mathcal{I}}(t,s) \Big) \\ =&\frac{1}{2}C_{LM}(t)C_{NO}(t)\Big(C^{-1}_{LM}(s)C^{-1}_{NO}(s)- C^{-1}_{LO}(s)C^{-1}_{MN}(s)\Big) 	
	\end{split}
\end{equation}
Third Invariant
\begin{equation}
	\begin{split}
		I_3(t,s)=\det(C(t,s))=\frac{J^2(t)}{J^2(s)}
	\end{split}
\end{equation}
Left Cauchy-Green Inverse
\begin{equation}
	\begin{split}
		b^{-1}_{ij}(t,s)=C_{MN}(s)F^{-1}_{Mi}(t)F^{-1}_{Nj}(t)
	\end{split}
\end{equation}

\section{Transient Network Tangents} \label{section:Tangents}
The tangents can be treated in an analogous way to the stress. The spatial tangent from all induced networks is given by
\begin{equation}
	\begin{split}
		\boldsymbol{h}^{**}(t) &= \intu_{0}^{t}  \gamma(t,s)\boldsymbol{h}^*(t,s) ds \, ,
	\end{split}
\end{equation}
where the spatial tangent of a particular network permits the decomposition
\begin{equation}
	\begin{split}
		\boldsymbol{h}^{*}(t,s) &= \sum_{i}\boldsymbol{\mathbb{A}}^{i}(s) : \boldsymbol{\mathbb{B}}^{i}(t) \, .
	\end{split}
\end{equation}
Then the spatial tangent may be treated (similar to the stress) as
\begin{equation} \label{TangDef}
	\boldsymbol{h}^{**}(t) = \sum_{i}\boldsymbol{\mathcal{G}}^{i}(t) : \boldsymbol{\mathbb{B}}^{i}(t)
\end{equation}
\begin{equation}
	\boldsymbol{\mathcal{G}}^{i}(t) =\intu_{0}^{t}  \gamma(t,s)\boldsymbol{\mathbb{A}}^{i}(s)ds \, ,
\end{equation}
for which the tangent history variables $\boldsymbol{\mathcal{G}}^{i}(t)$ are integrated in a consistent way with the stress tensor. For the invariant-based material models, the spatial tangent and their corresponding decompositions are given in Tables \ref{CNHTangenttable}\ref{BKTangenttable}, and \ref{YeohTangenttable}.

For the stretch-based model, the spatial tangent of the induced networks can be expressed through the Piola pushforward given by Equation \ref{Piola_push2}, 
where the material tangent of the induced networks is given by
\begin{equation}
	\boldsymbol{H}^{**}(t) = \intu_{0}^{t}  \gamma(t,s)\boldsymbol{H}^*(t,s)  ds ; \quad 	\boldsymbol{H}^*(t,s) = 4J(s)\frac{\partial^2 W(\boldsymbol{U}^2(t),\boldsymbol{U}^2(s))}{\partial \boldsymbol{C}(t)\partial \boldsymbol{C}(t)} \, .
\end{equation}
A decomposition is then sought of the form 
$\boldsymbol{H}^*(t,s)  =\sum_{i}\boldsymbol{\mathbb{A}}^{i}(s) : \boldsymbol{\mathbb{B}^*}^{i}(t)$, so that the spatial tangent is given by Equation \ref{TangDef}, with $\boldsymbol{\mathbb{B}}^{i}(t) =\frac{1}{J(t)}\boldsymbol{F}(t)\boldsymbol{F}(t) \boldsymbol{\mathbb{B}^*}^{i}(t) \boldsymbol{F}^{T}(t)\boldsymbol{F}^{T}(t)$.
For the term associated with the trace of a particular stretch measure, we have 
\begin{equation}
	H^{*}_{IJKL}(t,s)=  \mu  J(s)  U_{MN}^{-2\alpha}(s) L^{\alpha}_{MNIJKL}(t) \, ,
\end{equation}
where
\begin{equation}
  L^{\alpha}_{MNIJKL}:= 4 \frac{\partial^2 E^{\alpha}_{MN}}{\partial C_{IJ}\partial C_{KL}} \, .
\end{equation}
For the case of distinct eigenvalues, $\boldsymbol{L}^{\alpha}$ can be written as \cite{beex2019fusing,liu2024continuum}
\begin{equation} 
	\begin{split}
			\boldsymbol{L}^{\alpha}=&  \sum_{a=1}^{3} f_a \boldsymbol{M}_a \smallotimes \boldsymbol{M}_a \smallotimes \boldsymbol{M}_a +\sum_{a=1}^{3}\sum_{b\ne a}^{3} \xi_{ab}\big(\boldsymbol{G}^{**}_{abb}+\boldsymbol{G}^{**}_{bab}+\boldsymbol{G}^{**}_{bba}  \big) \\
			&+  \sum_{a=1}^{3}\sum_{b\ne a}^{3} \sum_{
			\substack{c\ne a \\ c\ne b }}^{3} \eta \boldsymbol{G}^{**}_{abc}
	\end{split}
\end{equation}
where
\begin{equation}
	\begin{split}
			\boldsymbol{G}^{**}_{abc}&=   \boldsymbol{N}_a  \smallotimes  \boldsymbol{N}_b \smallotimes  \boldsymbol{N}_c \smallotimes  \boldsymbol{N}_a  \smallotimes  \boldsymbol{N}_b \smallotimes  \boldsymbol{N}_c + \boldsymbol{N}_a \smallotimes  \boldsymbol{N}_b \smallotimes  \boldsymbol{N}_c \smallotimes  \boldsymbol{N}_a  \smallotimes  \boldsymbol{N}_c \smallotimes  \boldsymbol{N}_b  \\
			&+    \boldsymbol{N}_a  \smallotimes  \boldsymbol{N}_b \smallotimes  \boldsymbol{N}_a \smallotimes  \boldsymbol{N}_c  \smallotimes  \boldsymbol{N}_b \smallotimes  \boldsymbol{N}_c + \boldsymbol{N}_a \smallotimes  \boldsymbol{N}_b \smallotimes  \boldsymbol{N}_a \smallotimes  \boldsymbol{N}_c  \smallotimes  \boldsymbol{N}_c \smallotimes  \boldsymbol{N}_b  \\
			&+    \boldsymbol{N}_b  \smallotimes  \boldsymbol{N}_a \smallotimes  \boldsymbol{N}_c \smallotimes  \boldsymbol{N}_a  \smallotimes  \boldsymbol{N}_b \smallotimes  \boldsymbol{N}_c + \boldsymbol{N}_b \smallotimes  \boldsymbol{N}_a \smallotimes  \boldsymbol{N}_c \smallotimes  \boldsymbol{N}_a  \smallotimes  \boldsymbol{N}_c \smallotimes  \boldsymbol{N}_b  \\
			&+    \boldsymbol{N}_b  \smallotimes  \boldsymbol{N}_a \smallotimes  \boldsymbol{N}_a \smallotimes  \boldsymbol{N}_c  \smallotimes  \boldsymbol{N}_b \smallotimes  \boldsymbol{N}_c + \boldsymbol{N}_b \smallotimes  \boldsymbol{N}_a \smallotimes  \boldsymbol{N}_a \smallotimes  \boldsymbol{N}_c  \smallotimes  \boldsymbol{N}_c \smallotimes  \boldsymbol{N}_b  \\
	\end{split}
\end{equation}
\begin{equation}
	f_a := 4 \frac{\partial^2 E_a^{\alpha}}{\partial \lambda_a^2 \partial \lambda_a^2}= 2(\alpha-1) \lambda_a^{2(\alpha-2)}
\end{equation}
\begin{equation}
	\xi_{ab}:= \frac{ v_{ab}-\frac{1}{2}d_b}{(\lambda_a^2-\lambda_b^2)} 
\end{equation}
\begin{equation}
	\eta=\sum_{a=1}^{3}\sum_{b\ne a}^{3}\sum_{
		\substack{c\ne a \\ c\ne b }}^{3}  \frac{E_a^{\alpha}(\lambda_a)}{2(\lambda_a^2-\lambda_b^2)(\lambda_a^2-\lambda_c^2)} =\sum_{a=1}^{3}\sum_{b\ne a}^{3}\sum_{
		\substack{c\ne a \\ c\ne b }}^{3}  \frac{\lambda_a^{2\alpha}-1}{4\alpha(\lambda_a^2-\lambda_b^2)(\lambda_a^2-\lambda_c^2)} \, .
\end{equation}
In the case of repeated eigenvalues

\begin{equation}
	\lim\limits_{\lambda_b \rightarrow \lambda_a} \xi_{ab}=\frac{1}{8}f_a \, ; \quad \quad
	\lim\limits_{
		\substack{\lambda_b \rightarrow \lambda_a \\ \lambda_c \ne \lambda_a }} \eta=\xi_{ca} \, ; \quad \quad \lim\limits_{
		\substack{\lambda_b \rightarrow \lambda_a \\ \lambda_c \rightarrow \lambda_a }} \eta=\frac{1}{8}f_{a} \, .
\end{equation}
The decomposition for the tangent associated with the Ogden-Hill model is given in Table \ref{OgdenTangDecomp}. We note that for each material, there are no tangent history variables that have not already been computed as part of the Cauchy stress.

\begin{table}
	\centering
	\begin{tabular}{ |r |l| } 
		\hline
				$H_{ijkl}$ &$ \frac{1}{J} (\mu-\lambda \ln(J)) \Big(\delta_{ik}\delta_{jl} +\delta_{jk}\delta_{il}\Big) +\frac{\lambda}{J}  \delta_{ij}\delta_{kl}$ \\
		\hline
		$\mathbb{A}^{1}$ & $J(s) $  \\ 
		$\mathbb{A}^{2}$&$J(s)\ln(J(s))$ \\ 
		\hline
		$\mathbb{B}^{1}_{ijkl} $ & $ \frac{\lambda}{J(t)} \delta_{ij}\delta_{kl} +
\Big(\frac{\mu}{J(t)}-\frac{\lambda \ln(J(t))}{J(t)} \Big)  \Big(\delta_{ik}\delta_{jl} +\delta_{jk}\delta_{il}\Big) $   \\ 
		$\mathbb{B}^{2}_{ijkl}$&$ \frac{\lambda}{J(t)} \Big(\delta_{ik}\delta_{jl} +\delta_{jk}\delta_{il}\Big) $ \\
		\hline
	\end{tabular}
	\caption{Compressible neo-Hookean tangent decomposition}
	\label{CNHTangenttable}
\end{table}

\begin{table}
	\centering
	\begin{tabular}{ |r |l| } 
		\hline
		$H_{ijkl}$&$ \frac{1}{J}(1-f) \mu \Big(b^{-1}_{ik}\delta_{jl} +b^{-1}_{jk}\delta_{il} +b^{-1}_{il}\delta_{jk} +b^{-1}_{jl}\delta_{ik}  \Big)$ \\
		$\, $ & $+ \mu \beta \Big(  f J_3^{-\beta-1} + (1-f) J_3^{\beta-1} \Big) \delta_{kl} \delta_{ij}$  \\
		$\, $ & $+ \mu \Big( f J_3^{-\beta-1} - (1-f) J_3^{\beta-1} \Big)(\delta_{il}\delta_{jk}+ \delta_{jl}\delta_{ik} )$ \\
		\hline
		    	$\mathbb{A}^{1}_{MN} $  &$J(s)C_{MN}(s) $ \\
		$\mathbb{A}^{2}$ & $(J(s))^{1+\beta} $  \\
		$\mathbb{A}^{3}$  & $(J(s))^{1-\beta} $  \\ 
		\hline 		
		$\mathbb{B}^{1}_{MNijkl} $&$ \frac{1}{J(t)}(1-f) \mu \Big(F^{-1}_{Mi}(t)F^{-1}_{Nk}(t)\delta_{jl} +F^{-1}_{Mj}(t)F^{-1}_{Nk}(t)\delta_{il}      $    \\ 
$ \, $&$ \quad \quad \quad \quad \quad  +F^{-1}_{Mi}(t)F^{-1}_{Nl}(t)\delta_{jk} +F^{-1}_{Mj}(t)F^{-1}_{Nl}(t)\delta_{ik}  \Big)   $    \\ 
$\mathbb{B}^{2}_{ijkl}$&$\mu \beta f \big(J(t)\big)^{-\beta-1}  \delta_{kl} \delta_{ij} + \mu f \big(J(t)\big)^{-\beta-1} (\delta_{il}\delta_{jk}+ \delta_{jl}\delta_{ik} ) $
\\ 
$\mathbb{B}^{3}_{ijkl}$&$\mu \beta (1-f) \big(J(t)\big)^{\beta-1}  \delta_{kl} \delta_{ij} - \mu (1-f) \big(J(t)\big)^{\beta-1} (\delta_{il}\delta_{jk}+ \delta_{jl}\delta_{ik} ) $ \\
		\hline
	\end{tabular}
	\caption{Blatz-Ko tangent decomposition}
	\label{BKTangenttable}
\end{table}

\begin{table}
	\centering
	\begin{tabular}{ |r |l| } 
		\hline
		$H_{ijkl} $& $\frac{4c_1}{J}  \big( -\frac{1}{3}\bar{b}_{ij} \delta_{kl} - \frac{1}{3}\bar{b}_{kl} \delta_{ij} +  \frac{1}{9}\bar{I}_1\delta_{kl} \delta_{ij}+\frac{1}{6}\bar{I}_1(\delta_{ik}\delta_{jl}+\delta_{jk}\delta_{il})  \big)$\\
		$\,$ &$+\frac{4c_2}{J}  \big( 2  \bar{b}_{ij}\bar{b}_{kl} - \frac{4}{3}  \bar{I}_1 \bar{b}_{ij}\delta_{kl} -  \frac{4}{3}  \bar{I}_1 \delta_{ij}\bar{b}_{kl} +   \frac{4}{9} \bar{I}_1^2 \delta_{ij}\delta_{kl}   +\frac{1}{3}\bar{I}_1^2(\delta_{ik}\delta_{jl}+\delta_{jk}\delta_{il})          \big)$\\
		$\,$&$+\frac{4c_3}{J} \big( 6 \bar{I}_1  \bar{b}_{ij}\bar{b}_{kl} - 3 \bar{I}_1^2 \bar{b}_{ij}\delta_{kl} - 3  \bar{I}_1^2 \delta_{ij}\bar{b}_{kl}   +   \bar{I}_1^3  \delta_{ij}\delta_{kl}  +\frac{1}{2}\bar{I}_1^3 (\delta_{ik}\delta_{jl}+\delta_{jk}\delta_{il})      \big)  $\\
		\hline
					$\mathbb{A}^{1}_{MN}$  &$	J(s)\bar{C}_{MN}^{-1}(s)$   \\ 
		$ \mathbb{A}^{2}_{MNOP}$ & $ J(s)\bar{C}_{MN}^{-1}(s)\bar{C}_{OP}^{-1}(s)$  \\ 
		$ \mathbb{A}^{3}_{MNOPQR} $&$	J(s)\bar{C}_{MN}^{-1}(s)\bar{C}_{OP}^{-1}(s)\bar{C}_{QR}^{-1}(s) $ \\
		\hline
		$\mathbb{B}^{1}_{MNijkl} $ &$  \frac{4 c_1}{ J(t)} \Big(  -\frac{1}{3}\bar{F}_{iM}(t)\bar{F}_{jN}(t)\delta_{kl} - \frac{1}{3}\bar{F}_{kM}(t)\bar{F}_{lN}(t)\delta_{ij} $ \\
$\,$&$ \quad \quad \quad+ \frac{1}{9}\bar{C}_{MN}(t)\delta_{ij}\delta_{kl} + \frac{1}{6}\bar{C}_{MN}(t)\big(\delta_{ik}\delta_{jl}+\delta_{jk}\delta_{il}\big)    \Big)$    \\ 
$\mathbb{B}^{2}_{MNOPijkl}$ & $   \frac{4 c_2}{ J(t)} \Big(  2\bar{F}_{iM}(t)\bar{F}_{jN}(t)\bar{F}_{kO}(t)\bar{F}_{lP}(t)  $ \\
$\,$ & $ -\frac{4}{3} \bar{F}_{iM}(t)\bar{F}_{jN}(t)\delta_{kl}\bar{C}_{OP}(t)-\frac{4}{3} \bar{F}_{kM}(t)\bar{F}_{lN}(t)\delta_{ij}\bar{C}_{OP}(t)  $ \\
$\,$ & $	+\frac{4}{9}\bar{C}_{MN}(t)\bar{C}_{OP}(t)\delta_{ij}\delta_{kl} +\frac{1}{3}\bar{C}_{MN}(t)\bar{C}_{OP}(t)\big(\delta_{ik}\delta_{jl} + \delta_{jk}\delta_{il} \big) \Big)$ \\
$\mathbb{B}^{3}_{MNOPQRijkl} 	$&$	  \frac{4 c_3}{ J(t)} \Big(  6\bar{F}_{iM}(t)\bar{F}_{jN}(t)\bar{F}_{kO}(t)\bar{F}_{lP}(t)\bar{C}_{QR}(t)$ \\
$\,$ & $ -3 \bar{C}_{OP}(t)\bar{C}_{QR}(t) \big(  \bar{F}_{iM}(t)\bar{F}_{jN}(t)\delta_{kl}+\bar{F}_{kM}(t)\bar{F}_{lN}(t)\delta_{ij} \big) $  \\ 
$\,$&$ +\bar{C}_{MN}(t)\bar{C}_{OP}(t)\bar{C}_{QR}(t) \big( \delta_{ij}\delta_{kl}
+\frac{1}{2}(\delta_{ik}\delta_{jl} + \delta_{jk}\delta_{il} ) \big) \Big) $\\
		\hline 
	\end{tabular}
	\caption{Yeoh tangent decomposition}
	\label{YeohTangenttable}
\end{table}

\begin{table}
	\centering
	\begin{tabular}{ |r |l| } 
	\hline
		$\mathbb{A}^{\Tr}_{MN} $ &$J(s)  U_{MN}^{-2\alpha}(s)$   \\ 
		$\mathbb{A}^{\det} $ & $(J(s))^{2\alpha\beta+1} $  \\
			\hline
	$\mathbb{B}^{\Tr}_{MNijkl} $&$ \frac{\mu}{J(t)}  F_{iO}(t)F_{jP}(t)L^{\alpha}_{MNOPQR}(t)F_{kQ}(t)F_{lR}(t)$   \\ 
	$\mathbb{B}^{\det}_{ijkl}$&$ 2  \alpha\beta \mu  \big(J(t)\big)^{-2\alpha\beta-1}  \delta_{kl} \delta_{ij} + \mu  \big(J(t)\big)^{-2\alpha\beta-1} (\delta_{il}\delta_{jk}+ \delta_{jl}\delta_{ik} ) $ \\
		\hline 
	\end{tabular}
	\caption{Ogden-Hill tangent decomposition}
	\label{OgdenTangDecomp}
\end{table} 

\bibliographystyle{elsart-num-names}

\bibliography{Transient_Network}

\end{document}